\documentclass{cg_arxiv}

\usepackage[T1]{fontenc}
\usepackage{dfadobe}  
\usepackage{float}
\usepackage{afterpage}
\usepackage{placeins}
\usepackage{booktabs}
\usepackage{amssymb}
\usepackage{stackengine}
\usepackage{multirow}
\usepackage[colorlinks=true]{hyperref}

\usepackage{cite}  
\usepackage{xcolor}
\newif\ifshowchanges
\showchangesfalse 

\newcommand{\change}[1]{%
    \ifshowchanges
        \textcolor{red}{#1}%
    \else
        #1%
    \fi
}

\usepackage{pgfplots}
\usepackage{tikz}
\usepackage{orcidlink}

\usepackage{lineno}

\usepackage{graphicx}
\usepackage{amsmath}
\usepackage{color}
\usepackage{multirow}
\usepackage[normalem]{ulem}

\usepackage[labelfont={bf},textfont={it}]{caption}
\usepackage[font=it]{subcaption}
\usepackage[ruled,vlined,linesnumbered]{algorithm2e}

\SetCommentSty{mycommfont}
\SetKwComment{Comment}{$\triangleright$ }{}

\usepackage{bm}

\definecolor{mg5}{gray}{.5}
\definecolor{mg8}{gray}{.8}

\usepackage{import}

\newcommand{\ignorethis } [1] { }

\newcommand{\fignum     } [1] {\ref{#1}}

\newcommand{\fig        } [1] {Figure~\fignum{#1}}

\newcommand{\etal       }     {et~al.}

\usepackage{pict2e,picture}

\makeatletter
\DeclareRobustCommand{\Arrow}[1][]{%
\check@mathfonts
\if\relax\detokenize{#1}\relax
\settowidth{\dimen@}{$\m@th\rightarrow$}%
\else
\setlength{\dimen@}{#1}%
\fi
\sbox\z@{\usefont{U}{lasy}{m}{n}\symbol{41}}%
\begin{picture}(\dimen@,\ht\z@)
\roundcap
\put(\dimexpr\dimen@-.7\wd\z@,0){\usebox\z@}
\put(0,\fontdimen22\textfont2){\line(1,0){\dimen@}}
\end{picture}%
}
\makeatother

\title[Vector sketch animation generation with differentiable motion trajectories]
      {Vector sketch animation generation with \\ differentiable motion trajectories}

\author[Zhu \etal{}]{X. Zhu$^{1}$\orcidlink{0000-0002-3100-8481},  \,
X. Yang$^{1}$\orcidlink{0009-0004-0865-3783},  \,
S. Zheng$^{1}$,  \,
Z. Zhang$^{2}$\orcidlink{0009-0007-5597-504X},  \,
F. Gao$^{1}$,  \,
J. Huang$^{3}$ and\,
J. Chen\thanks{Corresponding author. Email: cjz@zjut.edu.cn.}$^{1}$\orcidlink{0000-0003-2780-6146} \\
{\parbox{\textwidth}{\centering $^1$Zhejiang University of Technology\, \,  $^2$ 	Hangzhou Dianzi University, \,\,  $^3$ 	Zhejiang Gongshang University}}
}

\begin{document}


\teaser{
    \centering
    \includegraphics[width=0.99\linewidth]{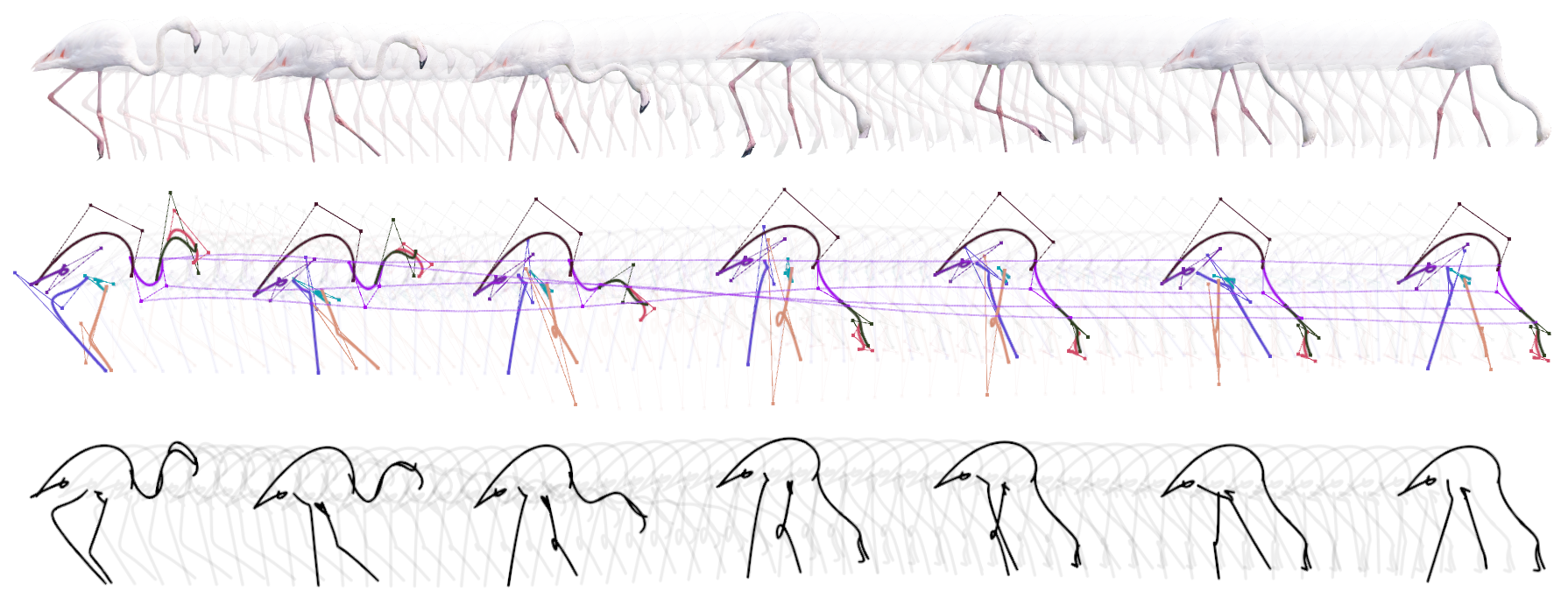}
    \caption{
        \change{Our method converts an object video (top) into a 2D vector sketch animation (bottom). We propose a differentiable motion trajectory with Bernstein basis to model cross-frame movement of stroke control points. The visualization in the middle displays their trajectories from a single stroke, and our approach boosts temporal/semantic consistency, performing robustly for sparse strokes and long videos.}
        \label{fig:teaser}
        \vspace{0.5em}
    }
}

\maketitle
\begin{abstract}
Sketching is a direct and inexpensive means of visual expression. Though image-based sketching has been well studied, video-based sketch animation generation is still very challenging due to the temporal coherence requirement. In this paper, we propose a novel end-to-end automatic generation approach for vector sketch animation. To solve the flickering issue, we introduce a Differentiable Motion Trajectory (DMT) representation that describes the frame-wise movement of stroke control points using differentiable polynomial-based trajectories. DMT enables global semantic gradient propagation across multiple frames, significantly improving the semantic consistency and temporal coherence, and producing high-framerate output. DMT employs a Bernstein basis to balance the sensitivity of polynomial parameters, thus achieving more stable optimization. Instead of implicit fields, we introduce sparse track points for explicit spatial modeling, which improves efficiency and supports long-duration video processing. Evaluations on DAVIS and LVOS datasets demonstrate the superiority of our approach over SOTA methods. Cross-domain validation on 3D models and text-to-video data confirms the robustness and compatibility of our approach.


\end{abstract}  


\section{Introduction} \label{sec:intro}
Vector animations, defined by geometric primitives (e.g., points, lines, curves, and polygons), offer significant advantages, including small file size, resolution-independent scalability, and ease of editing. Consequently, they are widely adopted in various applications, such as artistic design, industrial design, and data visualization~\cite{dalstein2015vector,bostock2011d3,tian2022survey,zhang2009vectorizing}. As a special form of vector animation, vector sketch animation simulates the human hand-drawing process with stylistic stroke feel, line dynamics, and visual abstraction, thus providing enhanced artistic expression and emotional appeal. This makes vector sketch animation particularly valuable for educational demonstrations, artistic communication, and industrial design~\cite{fan2023drawing,ha2017neural,aubert2014pleistocene,li2020differentiable}.
Vector sketch generation for static images has been well studied recently. CLIPasso~\cite{vinker2022clipasso} employs sets of Bézier curves to represent sketches, optimizing control point coordinates using CLIP perceptual loss~\cite{radford2021learning} and a differentiable rasterizer~\cite{li2020differentiable}. And CLIPascene\cite{vinker2023clipascene} further separates and optimizes foreground and background elements before blending them, achieving full-scene sketch generation. However, applying these methods for frame-by-frame sketch animation generation often results in unpleasant flickering~\cite{chen2023efficient}, as it neglects the temporal coherence of stroke topology across frames. To this end, Zheng et al.~\cite{zheng2024sketch} introduced a consistency loss based on Neural Layered Atlas (NLA)~\cite{kasten2021layered} to enhance cross-frame stroke stability; and Fang et al.~\cite{fang2025video} built upon CLIPascene and Zheng et al.'s work, employing Content Deformation Fields (CoDeF)~\cite{ouyang2024codef} to further improve temporal coherence in full-frame sketch animations. However, these methods have several limitations:
\begin{description}
    \item[Temporal popping and jitter] are still observed in their results, since the consistency losses employed by Zheng et al.~\cite{zheng2024sketch} and Fang et al.~\cite{fang2025video} can not strictly enforce continuity of stroke motion. Moreover, their ineffectiveness against minor temporal fluctuations makes jitter \change{between adjacent frames} an intractable issue.
    \item[Local semantics] are presented in strokes, because only nearby frames are considered during the optimization process. The weak association of strokes across frames loses the full leverage of the semantic information from the entire video.
    \item[Video length] is very limited, as their neural representations (e.g., deformation fields) fail to maintain spatial fidelity over long videos, resulting in matching errors and temporal incoherence in the generated sketch animation~\cite{fang2025video}.
\end{description}

To tackle these issues, a novel Differentiable Motion Trajectories (DMT) representation is proposed to enforce the temporal coherence of
strokes across frames. DMT represents vector sketch animations as a process driven by control points moving along continuous motion trajectories, thus \textbf{fully avoiding temporal popping and jitter}. Motion trajectories of control points are modeled as polynomials with Bernstein bases instead of the traditional power bases, enhancing learning stability. DMT enables the conversion of low-framerate input videos into high-framerate sketch animations, achieving video frame interpolation effects.


\textbf{To support long videos, an explicit representation of video spatial information via sparse tracking points is \change{adopted}}, instead of previous implicit representations using neural networks. These tracking points can not only be estimated from real videos using various computer vision techniques but also be precisely obtained from 3D model node projections when the input is a 3D animation, offering excellent editability and compatibility. The proposed approach was evaluated on two widely-used datasets: DAVIS~\cite{pont20172017} has short videos with 50-100 frames, and LVOS~\cite{hong2025lvos} has longer videos with more complex motions.



The main contributions of this paper are as follows:
\begin{itemize}
\item A differentiable motion trajectories representation that describes the frame-wise movement of stroke control points using differentiable polynomial-based trajectories is proposed. It significantly improves the temporal coherence and semantic consistency, and produces stable and high-framerate vector sketch animation.

\item An end-to-end framework for video-based vector sketch animation generation is presented. It incorporates a consistency loss based on sparse tracking, a stroke initialization strategy guided by video spatial information, and a differentiable rasterizer to produce sketch animations from long-duration videos.

\end{itemize}

\section{Previous work} \label{sec:previous}

\subsection{Vector sketch generation}
Existing style transfer techniques have been developed to generate sketch-style outputs from bitmap images~\cite{isola2017image,li2019photo,gryaditskaya2019opensketch,berger2013style}, along with pixel-level aligned approaches designed to directly generate vector sketches from bitmaps~\cite{carlier2020deepsvg,das2020beziersketch,reddy2021im2vec}. \change{However, the aforementioned methods suffer from the issue of stroke branching or folding.} To address the integration of vector graphics into deep learning pipelines, Li et al.~\cite{li2020differentiable} proposed a differentiable rendering framework for vector graphics that supports multiple geometric primitives, enabling the seamless incorporation of vector representations into deep models via a differentiable paradigm. Building upon this foundation, CLIPasso~\cite{vinker2022clipasso} leveraged a pre-trained CLIP model to compute perceptual loss between an input image and a Bézier curve-based generated output, optimizing curve control points through backpropagation to transform real-world images into abstract sketches. Expanding on this line of work, CLIPascene~\cite{vinker2023clipascene} decomposed input images into foreground and background regions for independent optimization, thereby generating full-frame sketches with tunable levels of abstraction. In contrast to image-driven methods, CLIPDraw~\cite{frans2022clipdraw} focused on text-to-vector-sketch generation, directly producing vector sketches from textual prompts. Recently, diffusion models~\cite{rombach2022high} have increasingly surpassed CLIP-based approaches for text-to-sketch tasks~\cite{xing2023diffsketcher, jain2023vectorfusion, xing2024svgdreamer}, with optimization guided by the Score Distillation Sampling (SDS) loss~\cite{poole2022dreamfusion} to enhance generation quality. Notably, SwiftSketch~\cite{arar2025swiftsketch} extended diffusion-based methods to image-conditioned vector sketch generation, replacing the time-consuming iterative optimization processes of prior works and enabling high-quality sketch production within seconds. Collectively, these advancements establish a solid foundation for exploring vector sketch animation generation—an area that remains an active and potential direction in computer graphics research.

\subsection{Vector sketch animation generation}
Extending image-based vector sketch generation to multi-frame animation poses the key challenge of maintaining temporal coherence of strokes across frames.
One category of methods relies on the spatial structure information from an input video to guide stroke optimization. Zheng et al.~\cite{zheng2024sketch} employed a Neural Layered Atlas (NLA)~\cite{kasten2021layered} to represent video spatial information as multi-plane images, constraining stroke positions via inter-frame point mapping. Fang et al.~\cite{fang2025video} utilized the more advanced Content Deformation Fields (CoDeF)~\cite{ouyang2024codef} to achieve stronger consistency. Liv3Stroke~\cite{lee2025recovering} extracted 3D point clouds from videos to guide the optimization of 3D curves. However, these methods struggle with long videos (hundreds of frames) due to limitations of NLA or CoDeF, leading to temporal incoherence. 

Another category of methods starts with a first-frame sketch and generates subsequent frames by adapting it. These approaches typically require an initial sketch and text prompts as input. Live-Sketch~\cite{gal2024breathing} and MoSketch~\cite{liu2025multi} first applied coarse-grained rigid transformations to the object in the first frame using an MLP, then refined the control points using SDS loss~\cite{poole2022dreamfusion}. However, this two-stage optimization struggles with complex motions, thus limiting its flexibility and ability to express complex motion semantics. GroupSketch~\cite{liang2025multi} involved users grouping sketch elements and setting keyframes; after generating motion trajectories via interpolation, it refined the motion using a Group-based Displacement Network (GDN). FlipSketch employed DDIM inversion~\cite{song2020denoising} to extract visual features from an input sketch, which were then fed into a Text-to-Video (T2V) diffusion model~\cite{wang2023modelscope} fine-tuned on data synthesized by Live-Sketch to generate raster animations. Rai et al.~\cite{rai2024enhancing} introduced a length-area regularization to enhance temporal coherence by estimating smooth motion of control points. The dependence on an initial sketch input, as well as the difficulty in representing complex motions, limits these methods to generating short animations or those with a limited range of motion.

\subsection{Animation driven by control points}
Control point-driven animation is a widely used technique in computer graphics that efficiently animates complex models or images through smooth deformations and motion by manipulating sparse control points~\cite{dalstein2015vector}. It is extensively applied in visual effects, game animation, vector illustration, and medical and scientific visualization~\cite{cui2015target,su2001control,ye2025vidanimator}. Traditional methods often rely on artists manually editing control points or extracting motion trajectories from videos, after which the computer automatically generates the deformation results~\cite{dalstein2015vector,su2018live,agarwala2004keyframe, bregler2002turning, guay2015space, santosa2013direct} .

With the advancement of deep learning, predicting motion trajectories and generating vivid animations without user intervention has become feasible~\cite{hinz2022charactergan,liu2019neuroskinning,jeruzalski2020nilbs}. Animation Drawing~\cite{smith2023method} utilized predefined character skeletal motions to animate characters drawn by children. AnaMoDiff~\cite{tanveer2024anamodiff} predicted optical flow fields from a reference video and applied the resulting deformations to an initial image. Siarohin et al.~\cite{siarohin2019first} generated image animations by applying local affine transformations to an initial image based on keypoints learned from videos. \change{AniClipart~\cite{wu2025aniclipart} and FlexiClip~\cite{khandelwal2025flexiclip} model control point trajectories via Bézier curves, leverage SDS for animated clipart generation, require the first frame as input, and optimize solely for motion trajectory smoothness without explicit semantic consistency constraints.} These approaches inspire us to explore control point-driven animation, which can represent complex motions with a compact set of parameters.

\section{Differential Motion Trajectories (DMT)} 

\label{sec:dmt}

This section introduces the theory of differentiable motion trajectories and their representations suitable for deep learning. 

\subsection{The DMT definition} 
The mathematical definition of differential motion trajectories is first introduced in this section. Taking video-to-sketch animation as an example, which involves converting a video of continuous motion into a sketch animation composed of 2D vector curves in each frame. Due to the temporal coherence requirement, vector curves appearing in consecutive frames should transform continuously.

Each frame in a video has $N_s$ strokes, one stroke can be represented as a Bézier curve with \(m+1\) control points \(\{P_0, P_1, ..., P_m\}\):
\begin{equation}
C(u) = \sum_{i=0}^{m} B_{m,i}(u) P_i
\end{equation}
where \(B_{m,i}(u) = \binom{m}{i} u^{i} (1-u)^{m-i}\) is the Bernstein basis function~\cite{farouki2012bernstein}.

The key insight is that any continuously varying Bézier curve can be approximated by a dynamic Bézier curve whose trajectories of control points $\{P_i\}$ are polynomial functions of time \(t\). We call trajectories of these control points as \textbf{Motion Trajectory}. As proven in Appendix~\ref{app:feasibility_proof}, if a Bézier curve changes continuously over time, then its control points must also vary continuously, i.e. its corresponding motion trajectories are continuous functions. By the Weierstrass approximation theorem~\cite{stone1948generalized}, these motion trajectories can be approximated well by polynomials with sufficient degrees. Thus, it is straightforward to represent a motion trajectory of a control point \(P\) as a polynomial:
\begin{equation}
P(t) = \sum_{i=0}^{n} t^i k_i
\label{equ:pt}
\end{equation}

\change{Integrating DMT into the automatic sketch animation generation pipeline allows the semantic loss for a single frame to back-propagate not only to the control point parameters of that frame's strokes but also to the global parameters of DMT. \textbf{It encodes global video semantic information into strokes of each frame,} allowing to represent complete semantics with a small number of strokes. }

Since polynomial functions $P(t)$ are infinitely differentiable, we call trajectories of control points \textbf{Differential Motion Trajectories (DMT in short)}. The DMT continuity enforces the temporal coherence of strokes across frames. When DMT is used in the differentiable optimization process for 2D vector graphics, it allows gradients, which are originally passed to the coordinates of the control point, to be further back-propagated to the set of polynomial parameters \(\{k_i\}\). This mechanism enables local loss optimization on a single frame to influence the global state of the entire sketch animation. \textbf{It encodes global video semantic information into strokes of each frame,} ensuring that the generated vector graphics exhibit semantic consistency across frames and allowing to represent complete semantics with a small number of strokes.


\subsection{Bernstein basis representation} 

The parameter set \(\{k_i\}\) in Equation~\ref{equ:pt} can be optimized by machine learning methods. However, the power basis representation suffers from uneven sensitivity across the temporal domain, which adversely affects gradient-based optimization~\cite{goodfellow2016deep}.

The L1 norm of the sensitivity for the power basis form is:
\begin{equation}
\| S(t) \|_1 = \sum_{i=0}^{n} \left| \frac{\partial P(t)}{\partial k_i} \right| = \sum_{i=0}^{n} t^i
\end{equation}

This sensitivity varies dramatically with \(t\). For example, \(\| S(0) \|_1 = 1\) when \(t=0\), while \(\| S(1) \|_1 = n+1\) when\(t=1\). This variation causes two optimization challenges:

\begin{description}
    \item[Gradient vanishing:] When \(t\) is small, the sensitivity approaches 1, resulting in minimal gradient magnitudes that hinder parameter updates for early frames.
    \item[Gradient explosion:] When \(t\) is large, the sensitivity grows exponentially, causing overlarge gradients and training instability.
\end{description}

To address these issues, a Bernstein basis representation is adopted~\cite{zhang2015level,seeger2004gaussian}:
\begin{equation}
P'(t) = \sum_{i=0}^{n} B_{n,i}(t) \cdot q_i
\end{equation}
\vspace{-10pt}

It achieves uniform sensitivity across the temporal domain:
\begin{equation}
\| S'(t) \|_1 = \sum_{i=0}^{n} \left| \frac{\partial P'(t)}{\partial q_i} \right| = \sum_{i=0}^{n} B_{n,i}(t) = 1
\end{equation}

\vspace{-5pt}

This constant sensitivity ensures stable gradient flow throughout optimization, mitigating the risks of vanishing/exploding gradients and enabling more reliable convergence~\cite{hastie2009elements,de1978practical}. A corresponding ablation study can be found in Section~\ref{subsubsec:basis_comparison}.

\begin{figure*}[!t]
    \centering
    \includegraphics[width=0.94\linewidth]{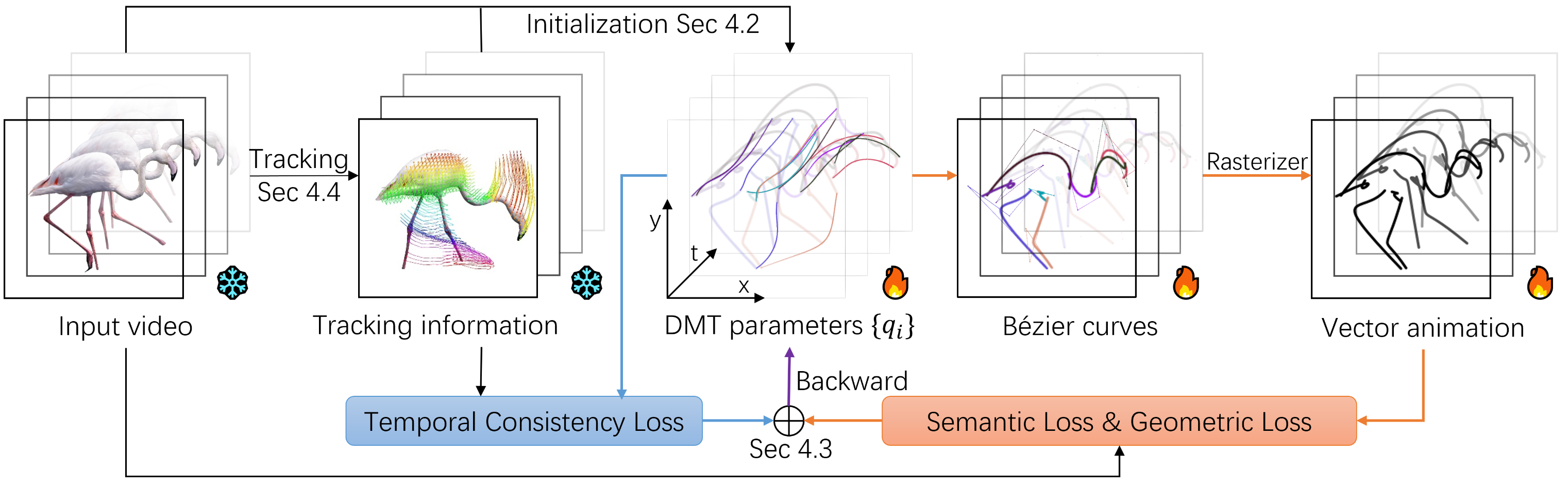}
    \caption{The framework of our DMT-based vector sketch animation generation. First, we obtain the tracking information from the video and initialize the DMT parameters. Then, these parameters are iteratively optimized to make the rasterized sketch animation semantically and geometrically close to the input video, and the stroke movement trajectories are consistent with the tracking information.}
    \label{fig:pipeline}
\end{figure*}

\section{Methodology} \label{sec:method}


\subsection{Overview}

\fig{fig:pipeline} shows the framework of our vector sketch animation generation method based on Differentiable Motion Trajectories (DMT). It first extracts tracking information of sparsely sampled points from the input video using computer vision algorithms or from 3D animations using graphic rendering. Subsequently, parameters of DMT corresponding to the vector strokes and their control points are initialized.

During the iterative optimization process, a differentiable rasterizer~\cite{li2020differentiable} is used to render vector graphics into images frame by frame, and the semantic loss and the geometric loss between the generated results and the original video frames are computed using the CLIP model~\cite{radford2021learning}. And the temporal consistency loss is constructed to leverage the video's spatial information. By back-propagating losses to the motion trajectory parameters, trajectory functions are jointly optimized, ultimately producing a sketch animation with visual coherence and semantic consistency.

By representing the vector animation as the evolution of Bézier curve control points along continuous motion trajectories, the generated results are naturally continuous between frames, fundamentally avoiding jitter and flickering artifacts. During optimization, the semantic loss from each frame can be back-propagated to the global motion trajectory parameters, enabling each frame's strokes to encapsulate the semantic information of the entire video. Furthermore, the introduced consistency loss constrains the position of the same stroke across various frames of the video, aligning it with the original video content, thereby enhancing the temporal semantic consistency and spatial stability of the strokes.

\subsection{Initialization}
\paragraph*{Motion-aware probability density map}. CLIPasso~\cite{vinker2022clipasso} obtains initial coordinates of strokes by randomly sampling the attention map, which is suitable for static images. However, for dynamic videos, this approach often fails to allocate enough strokes to depict moving regions. Therefore, motion magnitude must be considered besides semantic attention and integrated into the probability density map. To this end, a motion weight for the \( j \)-th sample point is first computed:
\begin{equation}
V_m(j) = \left( \sum_{i=1}^{N_f-1} \| {TRACK}_{i,j} - {TRACK}_{i-1,j} \| \right)^{1/2}
\end{equation}
\vspace{-5pt}

\noindent where \( N_f \) is the number of video frames, and \( {TRACK}_{i,j} \) denotes the 2D coordinates of the \( j \)-th tracked sample point in the \( i \)-th frame. These tracks can be obtained via estimation from video or accurately acquired from 3D animation (see Section~\ref{sec:tracking} for details). 

\( V_m(j) \) is defined on sparse tracking positions, a radial basis function interpolation is then employed to obtain a per-pixel motion weight across the entire image, which is further normalized to the range [0, 1] to form the motion heatmap \( M_{{motion}} \). It first blends with the CLIP attention map $M_{{attention}}$ linearly, and then multiplies with an XDoG edge map~\cite{winnemoller2012xdog} to encourage initial stroke placement near object contours. The final probability density map is:

\vspace{-8pt}
\begin{equation}
M = M_{{XDoG}} \otimes \big( (1 - \beta) \cdot M_{{attention}} + \beta \cdot M_{{motion}} \big)
\end{equation}
\vspace{-10pt}

\noindent where \( \otimes \) is the Hadamard product, and \( \beta = 0.5 \) in all experiments. 

Initial stroke coordinates are sampled according to this probability density distribution, favoring regions with higher density~\cite{fisher1922mathematical}. An ablation study in Section~\ref{subsubsec:motion_heatmaps} compares results with and without the motion heatmap.

\paragraph*{The DMT initialization.}
\change{The degree of the time polynomial $n$ is set to $N_f/4$ as the default. Users can adjust this parameter based on the smoothness of the target motion.} To avoid local minima and accelerate convergence, we initialize strokes such that their trajectories roughly approximate the true object motion. Instead of randomly selecting pixel positions as in CLIPasso, a set of points (equal to the number of strokes) is randomly sampled from the sparsely tracked points with motion trajectories. These sampled trajectories serve as fitting targets for the initial strokes.

For each motion trajectory, a polynomial should be fitted with a Bernstein basis. Three fitting methods are evaluated: polynomial interpolation, least squares, and ridge regression. Considering the fitting error and numerical stability, the ridge regression is ultimately adopt for polynomial fitting. A comparison with these three methods can be found in Appendix~\ref{sec:appendix}.

\paragraph*{Stroke width calculation.}
Inspired by SketchVideo~\cite{zheng2024sketch}, the stroke width per frame should be adapted to the relative size of the target object to prevent artifacts. When the object is small, thinner strokes are used to avoid \change{unexpected overlap of thick strokes in dense areas}; when the object is large, thicker strokes are employed to minimize blank areas.

The stroke width for the \( i \)-th frame, \( {width}_i \), is adapted according to the area of the object mask (generated by a UNet2 network~\cite{qin2020u2}) relative to the image area:
\begin{equation}
{width}_i = {W}_{max} \cdot \sqrt{{Area}_i/(W \cdot H)}
\end{equation}

\vspace{-5pt}

\noindent where \change{\( {W}_{max}=3 \)} denotes the maximum stroke width, \( {Area}_i \) is the pixel count of the target object in the mask for the \( i \)-th frame, and \( W \) and \( H \) are the image width and height respectively.

\subsection{Loss functions}

\change{We propose a comprehensive loss function: the semantic and geometric losses are derived from CLIPasso~\cite{vinker2022clipasso}, while a temporal constraint is newly added to our design.} The total loss is defined as:
\begin{equation}
\mathcal{L}_{{total}} = w_s \cdot \mathcal{L}_{{sem}} + w_g \cdot \mathcal{L}_{{geo}} + w_c \cdot \mathcal{L}_{{cons}}
\end{equation}

\noindent\change{where $w_s,w_g$ and $w_c$ are weighting coefficients. In optimization, they are set to 0.1,1 and 2, respectively.}

\noindent \textbf{Semantic loss.}
The final layer of the CLIP encoder captures high-level semantic information. The semantic loss $\mathcal{L}_{{sem}}$ is defined as the cosine-based distance between the CLIP embeddings of the rasterized vector graphics and the input video frames:

\vspace{-15pt}

\begin{equation}
\mathcal{L}_{{sem}} = \sum_{i=0}^{N_f-1} {dist}\bigg( {CLIP}\Big(R\big(\mathbb{C}(i)\big)\Big), {CLIP}(I_i) \bigg)
\end{equation}

\vspace{-5pt}

\noindent where $\mathbb{C}(i)$ denotes the set of Bézier curves at frame $i$, $R(\mathbb{C}(i))$ is the image rendered by the differentiable rasterizer, $I_i$ is the $i$-th input video frame, and ${dist}(x,y) = 1 - \frac{x \cdot y}{\|x\|\|y\|}$ computes the cosine-based distance.

\noindent \textbf{Geometric Loss.}
Intermediate layers of the CLIP network contain more spatial and geometric information compared to the final layer. Thus, a geometric loss $\mathcal{L}_{{geo}}$ is defined as the L2 distance between intermediate feature activations:
\begin{equation}
\mathcal{L}_{{geo}} = \sum_{i=0}^{N_f-1} \sum_{l \in \{3,4\}} \Big|\Big| {CLIP}_l\Big(R\big(\mathbb{C}(i)\big)\Big) - {CLIP}_l(I_i) \Big|\Big|^2
\end{equation}
\noindent where ${CLIP}_l$ is the CLIP encoder activation at layer $l$. 

Both $\mathcal{L}_{{sem}}$ and $\mathcal{L}_{{geo}}$ are derived from the CLIP model and can be computed efficiently in a single forward pass (implementation details in Section~\ref{sssec:memory_opt}).

\noindent \textbf{Temporal consistency loss.}
To ensure stroke motion follows the underlying video motion, a temporal consistency loss is introduced, which is the consistency sum of all strokes in all frames:

\vspace{-15pt}

\begin{equation}
\mathcal{L}_{{cons}} = \sum_{i=0}^{N_f-1} \sum_{j=1}^{N_s} \mathcal{L}_{{cons}}^{{stroke}}(i,j)
\end{equation}

\vspace{-5pt}

\noindent where $N_s$ is the number of strokes, and $\mathcal{L}_{{cons}}^{{stroke}}(i,j)$ measures the consistency for the $j$-th stroke at the $i$-th frame.

For computational efficiency, stroke consistency is computed by uniformly sampling $N_p$ points along each Bézier curve:

\vspace{-15pt}

\begin{equation}
\mathcal{L}_{{cons}}^{{stroke}}(i,j) = \frac{1}{N_p} \sum_{k=0}^{N_p-1} \mathcal{L}_{{cons}}^{{point}}\big(i,j,u=k/\left(N_p-1\right)\big)
\end{equation}

\vspace{-5pt}

The point-wise consistency loss $\mathcal{L}_{{cons}}^{{point}}(i,j,u)$ measures the average L2 distance between the position of the sampled point across all frames and its corresponding position along the tracked trajectory:

\vspace{-15pt}

\begin{equation}
\mathcal{L}_{{cons}}^{{point}}(i,j,u) = \frac{1}{N_f} \sum_{t=0}^{N_f-1} \| \mathbf{T}(\mathcal{C}(i,j,u), i, t) - \mathcal{C}(t,j,u) \|^2
\end{equation}

\vspace{-5pt}

\noindent where $\mathcal{C}(i,j,u)$ represents the 2D coordinates of the $j$-th Bézier curve at parameter $u$ in frame $i$, and $\mathbf{T}(\mathbf{p}, i, t)$ predicts the position of pixel $\mathbf{p}$ from frame $i$ to frame $t$ using the video motion trajectory. Since we only have sparse trajectory data, an interpolation scheme is employed, whose details can be found in Section~\ref{sssec:sparse_to_dense}.

\subsection{Tracking information}\label{sec:tracking}
Existing approaches typically employ neural-based methods such as Neural Layered Atlas (NLA)~\cite{kasten2021layered} or Content Deformation Fields (CoDeF)~\cite{ouyang2024codef} to maintain temporal consistency of strokes. However, these methods suffer from two main limitations: 1) These methods are difficult to processing long videos accurately and will increasing the inconsistency when the number of frames is over 100.~\cite{fang2025video}; 2) the encoded information is embedded within neural network weights, lacking interpretability and editability, which restricts practical application flexibility.

\begin{figure*}[!t]
    \centering
    \begin{tabular}{cccccccc}
    \setlength{\tabcolsep}{0pt} 
    
    \raisebox{-.5\height}{\includegraphics[width=0.1\linewidth]{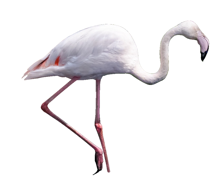}} &
    \raisebox{-.5\height}{\includegraphics[width=0.1\linewidth]{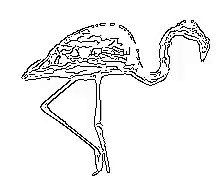}} &
    \raisebox{-.5\height}{\includegraphics[width=0.1\linewidth]{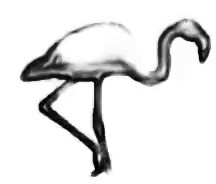}} &
    \raisebox{-.5\height}{\includegraphics[width=0.1\linewidth]{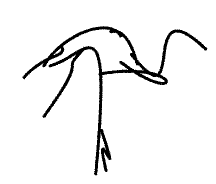}} &
    \raisebox{-.5\height}{\includegraphics[width=0.1\linewidth]{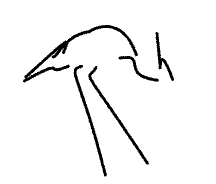}} &
    \raisebox{-.5\height}{\includegraphics[width=0.1\linewidth]{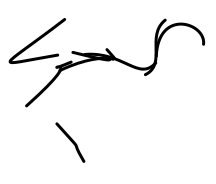}} &
    \raisebox{-.5\height}{\includegraphics[width=0.1\linewidth]{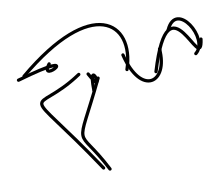}} &
    \raisebox{-.5\height}{\includegraphics[width=0.1\linewidth]{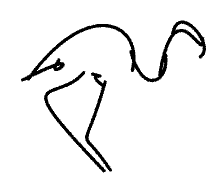}} \\

    \raisebox{-.5\height}{\includegraphics[width=0.1\linewidth]{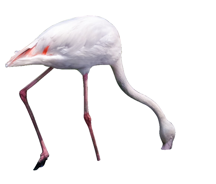}} &
    \raisebox{-.5\height}{\includegraphics[width=0.1\linewidth]{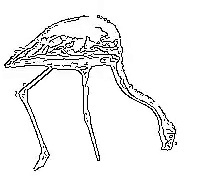}} &
    \raisebox{-.5\height}{\includegraphics[width=0.1\linewidth]{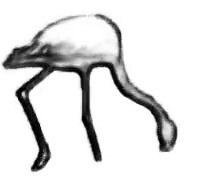}} &
    \raisebox{-.5\height}{\includegraphics[width=0.1\linewidth]{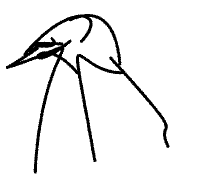}} &
    \raisebox{-.5\height}{\includegraphics[width=0.1\linewidth]{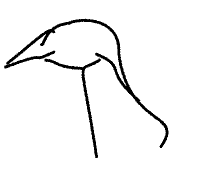}} &
    \raisebox{-.5\height}{\includegraphics[width=0.1\linewidth]{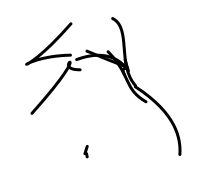}} &
    \raisebox{-.5\height}{\includegraphics[width=0.1\linewidth]{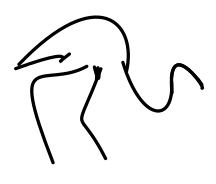}} &
    \raisebox{-.5\height}{\includegraphics[width=0.1\linewidth]{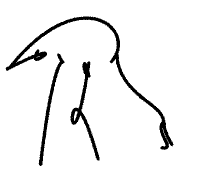}} \\

    \raisebox{-.5\height}{\includegraphics[width=0.1\linewidth]{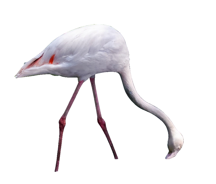}} &
    \raisebox{-.5\height}{\includegraphics[width=0.1\linewidth]{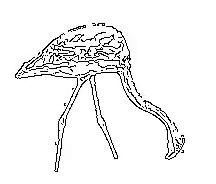}} &
    \raisebox{-.5\height}{\includegraphics[width=0.1\linewidth]{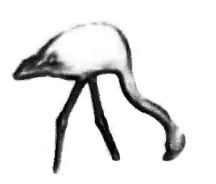}} &
    \raisebox{-.5\height}{\includegraphics[width=0.1\linewidth]{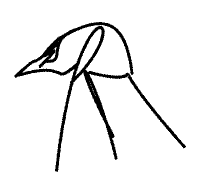}} &
    \raisebox{-.5\height}{\includegraphics[width=0.1\linewidth]{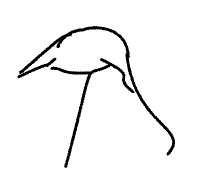}} &
    \raisebox{-.5\height}{\includegraphics[width=0.1\linewidth]{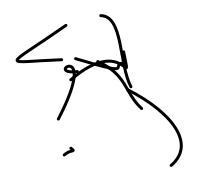}} &
    \raisebox{-.5\height}{\includegraphics[width=0.1\linewidth]{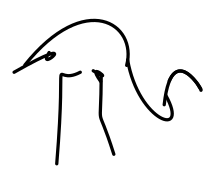}} &
    \raisebox{-.5\height}{\includegraphics[width=0.1\linewidth]{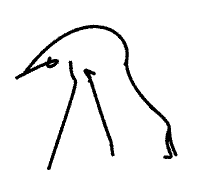}} \\

    Input & Canny & HED  & CLIPasso & SketchVideo  & \change{Fang et al.} & \change{LiveSketch-MLP} & Ours \\
     & \cite{canny2009computational} & \cite{xie2015holistically} & \cite{vinker2022clipasso} & \cite{zheng2024sketch} & \cite{fang2025video} & ~\cite{gal2024breathing} &  \\
    \end{tabular}
    \caption{\textbf{Comparisons with SOTA methods with 3 frames (the first, 25th, and 49th frames from top to bottom).} Canny and HED do not produce vector graphics. CLIPasso, SketchVideo, and Fang et al. produce competitive results, but suffer from the temporal incoherence issue, such as legs and the tail. \change{LiveSketch-MLP fails to capture the bent posture of the flamingo's head. The results produced by our approach achieve superior temporal coherence and semantic consistency.} }
    \label{fig:comparison-diff-methon}
\end{figure*}

To address these issues, an explicit representation of video spatial information based on sparse sample point trajectories is proposed. Our method uniformly samples a moderate number of feature points (typically 2,000 to 10,000) from the video and tracks their positions across frames. Temporal consistency of sketch strokes is enforced by measuring the similarity between a stroke's motion trajectory and those of nearby sampled points.

\noindent \textbf{Sparse tracking for natural videos. }
For natural videos, CoTracker~\cite{karaev2024cotracker3} is employed as our base tracker and enhances its memory management to support long video sequences. Implementation details are discussed in Section~\ref{sssec:memory_opt}.

\noindent \textbf{Sparse tracking for 3D animation. }
For 3D animation input, vertices on the model surface from each sequence frame are uniformly sampled, and their screen-space coordinates are recorded. To ensure strict data alignment, a customized rendering script is used to synchronously capture the frame and record vertex coordinates after each frame is rendered, avoiding inter-frame deviations caused by GPU-CPU asynchronous execution in real-time rendering. 

\subsection{Implementation details}

\subsubsection{Bernstein polynomials with high degrees}
\label{sssec:high_degree_bernstein}

Representing complex motion trajectories requires higher-degree polynomials. For Bernstein basis functions $B_{n,i}(t) = \binom{n}{i} t^i (1-t)^{n-i}$, when $n$ is large ($n > 100$), the binomial coefficient $\binom{n}{i}$ can become extremely large while $t^i$ or $(1-t)^{n-i}$ becomes extremely small, leading to numerical precision issues with PyTorch's default 32-bit floating-point arithmetic. While switching to 64-bit precision provides some relief, it becomes insufficient as $n$ increases further.

Considering that the value of $P_x(t)$ should not exceed the image dimensions and does not require high decimal precision, a logarithmic approach is employed. $\log\binom{n}{i}$, $\log(t^i)$, and $\log((1-t)^{n-i})$ are computed first, then sum them to obtain $\log(\binom{n}{i}t^i(1-t)^{n-i})$, and finally convert back to the non-logarithmic value. This logarithmic strategy maintains sufficient precision with 32-bit floating-point arithmetic while supporting much larger values of $n$.

\subsubsection{Memory optimization for long videos}
\label{sssec:memory_opt}
Long video generation and analysis are often constrained by memory limitations ~\cite{ouyang2024codef,yan2021videogpt,liu2022video}. To this end, a memory-efficient optimization strategy is designed for both video spatial information extraction and vector animation generation stages. The related experimental results are shown in Table~\ref{tab:performance}.

\begin{figure*}[!t]
    \centering
    \setlength{\tabcolsep}{0pt} 
    
    \begin{tabular}{cc|c c|c c|c c|c c|c c}
    \multirow{4}{*}{
        \begin{tikzpicture}
        [scale=1, baseline=(current bounding box.center)]
        \draw[->, thick, line width=1pt] (0,1) -- (0,-2);
        \node at (0,-2.3) {$t$};
    \end{tikzpicture}}
    &
    \raisebox{-.5\height}{\includegraphics[width=0.07\linewidth]{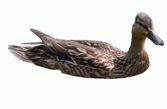}} &
    \raisebox{-.5\height}{\includegraphics[width=0.075\linewidth]{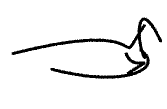}} &
    \raisebox{-.5\height}{\includegraphics[width=0.075\linewidth]{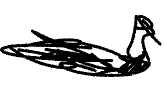}} &
    \raisebox{-.5\height}{\includegraphics[width=0.075\linewidth]{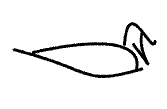}} &
    \raisebox{-.5\height}{\includegraphics[width=0.075\linewidth]{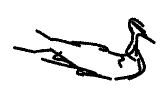}} &
    \raisebox{-.5\height}{\includegraphics[width=0.075\linewidth]{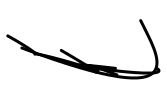}} &
    \raisebox{-.5\height}{\includegraphics[width=0.075\linewidth]{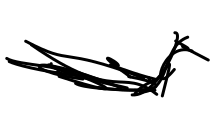}}&
    \raisebox{-.5\height}{\includegraphics[width=0.075\linewidth]{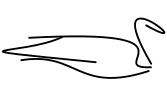}} &
    \raisebox{-.5\height}{\includegraphics[width=0.075\linewidth]{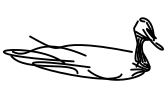}}&
    \raisebox{-.5\height}{\includegraphics[width=0.075\linewidth]{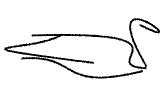}} &
    \raisebox{-.5\height}{\includegraphics[width=0.075\linewidth]{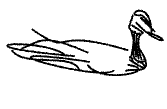}} \\
    
    &
    \raisebox{-.5\height}{\includegraphics[width=0.075\linewidth]{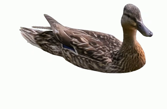}} &
    \raisebox{-.5\height}{\includegraphics[width=0.075\linewidth]{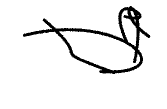}} &
    \raisebox{-.5\height}{\includegraphics[width=0.075\linewidth]{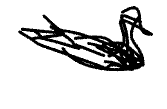}} &
    \raisebox{-.5\height}{\includegraphics[width=0.075\linewidth]{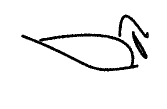}} &
    \raisebox{-.5\height}{\includegraphics[width=0.075\linewidth]{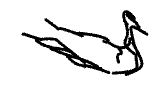}} &
    \raisebox{-.5\height}{\includegraphics[width=0.075\linewidth]{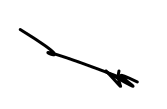}} &
    \raisebox{-.5\height}{\includegraphics[width=0.075\linewidth]{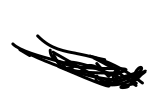}}&
    \raisebox{-.5\height}{\includegraphics[width=0.075\linewidth]{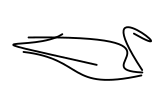}} &
    \raisebox{-.5\height}{\includegraphics[width=0.075\linewidth]{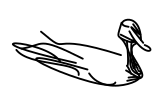}}&
    \raisebox{-.5\height}{\includegraphics[width=0.075\linewidth]{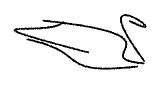}} &
    \raisebox{-.5\height}{\includegraphics[width=0.075\linewidth]{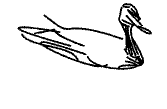}} \\
    
    &
    \raisebox{-.5\height}{\includegraphics[width=0.075\linewidth]{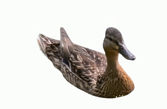}} &
    \raisebox{-.5\height}{\includegraphics[width=0.075\linewidth]{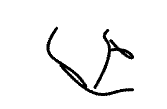}} &
    \raisebox{-.5\height}{\includegraphics[width=0.075\linewidth]{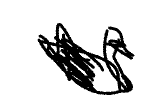}} &
    \raisebox{-.5\height}{\includegraphics[width=0.075\linewidth]{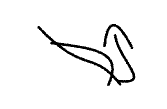}} &
    \raisebox{-.5\height}{\includegraphics[width=0.075\linewidth]{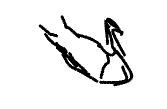}} &
    \raisebox{-.5\height}{\includegraphics[width=0.075\linewidth]{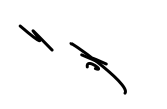}} &
    \raisebox{-.5\height}{\includegraphics[width=0.075\linewidth]{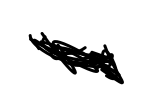}}&
    \raisebox{-.5\height}{\includegraphics[width=0.075\linewidth]{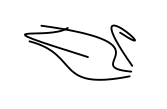}} &
    \raisebox{-.5\height}{\includegraphics[width=0.075\linewidth]{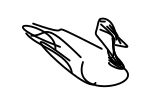}}&
    \raisebox{-.5\height}{\includegraphics[width=0.075\linewidth]{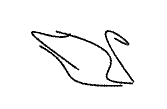}} &
    \raisebox{-.5\height}{\includegraphics[width=0.075\linewidth]{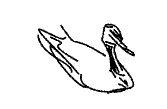}} \\
    
    &
    \raisebox{-.5\height}{\includegraphics[width=0.075\linewidth]{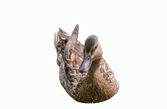}} &
    \raisebox{-.5\height}{\includegraphics[width=0.075\linewidth]{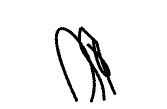}} &
    \raisebox{-.5\height}{\includegraphics[width=0.075\linewidth]{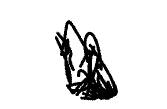}} &
    \raisebox{-.5\height}{\includegraphics[width=0.075\linewidth]{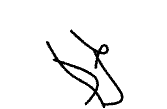}} &
    \raisebox{-.5\height}{\includegraphics[width=0.075\linewidth]{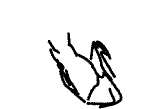}} &
    \raisebox{-.5\height}{\includegraphics[width=0.075\linewidth]{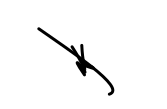}} &
    \raisebox{-.5\height}{\includegraphics[width=0.075\linewidth]{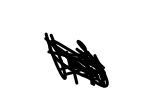}}&
    \raisebox{-.5\height}{\includegraphics[width=0.075\linewidth]{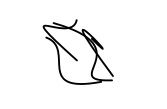}} &
    \raisebox{-.5\height}{\includegraphics[width=0.075\linewidth]{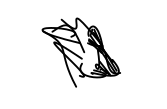}}&
    \raisebox{-.5\height}{\includegraphics[width=0.075\linewidth]{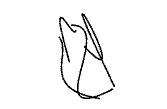}} &
    \raisebox{-.5\height}{\includegraphics[width=0.075\linewidth]{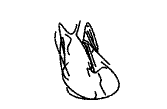}} \\
    
    & Input & 4 strokes & 16 strokes & 4 strokes& 16 strokes & \change{4 strokes} & \change{16 strokes} & \change{4 strokes} & \change{16 strokes} & 4 strokes & 16 strokes \\
    
    &
    \multicolumn{1}{c|}{} & 
    \multicolumn{2}{c|}{Clipasso~\cite{vinker2022clipasso}} & 
    \multicolumn{2}{c|}{SketchVideo~\cite{zheng2024sketch}} & 
    \multicolumn{2}{c|}{\change{Fang et al.}~\cite{fang2025video}} & 
    \multicolumn{2}{c|}{\change{LiveSketch-MLP}~\cite{gal2024breathing}}& 
    \multicolumn{2}{c}{Ours} \\
    \end{tabular}
    
    \caption{\change{\textbf{Comparisons of 5 SOTA methods with different numbers of strokes.}  CLIPasso lacks temporal coherence; CLIPasso, SketchVideo and Fang et al. suffer from semantic loss with a small number of strokes. LiveSketch-MLP shows less cross-frame semantic consistency, particularly with increasing frame index \( t \).
    }}
    \label{fig:comparison-diff-pathnum}
\end{figure*}

\begin{figure*}[t]
    \centering
    \begin{subfigure}[b]{\textwidth}
        \centering
        \includegraphics[width=0.1\textwidth]{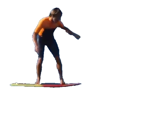}\hfill
        \includegraphics[width=0.1\textwidth]{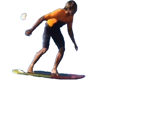}\hfill
        \includegraphics[width=0.1\textwidth]{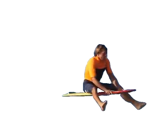}\hfill
        \includegraphics[width=0.1\textwidth]{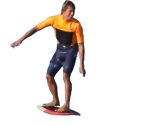}\hfill
        \includegraphics[width=0.1\textwidth]{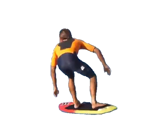}\hfill
        \includegraphics[width=0.1\textwidth]{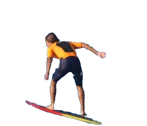}\hfill
        \includegraphics[width=0.1\textwidth]{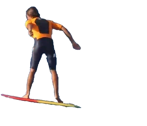}\hfill
        \includegraphics[width=0.1\textwidth]{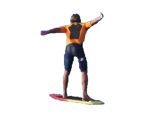}\hfill
        \includegraphics[width=0.1\textwidth]{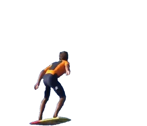}
        
        \stackunder{\includegraphics[width=0.1\textwidth]{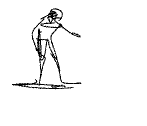}}{\small \# 0}\hfill
        \stackunder{\includegraphics[width=0.1\textwidth]{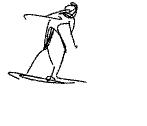}}{\small \# 120}\hfill
        \stackunder{\includegraphics[width=0.1\textwidth]{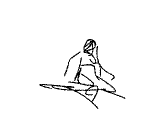}}{\small \# 240}\hfill
        \stackunder{\includegraphics[width=0.1\textwidth]{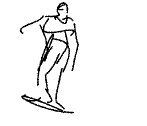}}{\small \# 360}\hfill
        \stackunder{\includegraphics[width=0.1\textwidth]{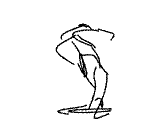}}{\small \# 480}\hfill
        \stackunder{\includegraphics[width=0.1\textwidth]{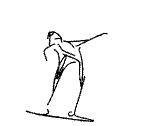}}{\small \# 600}\hfill
        \stackunder{\includegraphics[width=0.1\textwidth]{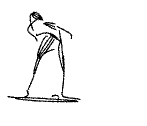}}{\small \# 720}\hfill
        \stackunder{\includegraphics[width=0.1\textwidth]{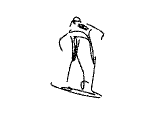}}{\small \# 840}\hfill
        \stackunder{\includegraphics[width=0.1\textwidth]{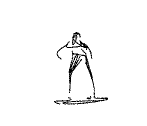}}{\small \# 1080}
        
    \end{subfigure}
    \caption{\change{The result of \textbf{long} Surfing video~\cite{hong2025lvos}. (6FPS input with 300 frames; 24FPS output with 1200 frames)}}
    \label{fig:long_video_results}
\end{figure*}

\noindent \textbf{CoTracker inference optimization.}
The original CoTracker~\cite{karaev2024cotracker3} maintains the state of all frames and all tracked points in GPU memory, leading to linear memory growth with sequence length, and thus can only process videos with fewer than 100 frames. To overcome this limitation, the tracking pipeline is redesigned to store intermediate features and inactive point cloud data in host memory (CPU RAM), transfer data to GPU only during essential computations such as optical flow propagation and feature matching. This heterogeneous storage strategy significantly reduces peak GPU memory usage, enabling sparse tracking for videos with more than 800 frames.

\noindent \textbf{Multi-frame gradient accumulation optimization.}
In gradient-based video optimization, the total loss depends on accumulated errors across all frames. Direct backpropagation leads to memory usage proportional to the number of frames. Profiling reveals that CLIP-based losses (e.g., $\mathcal{L}_{{geo}}$ and $\mathcal{L}_{{sem}}$) constitute the primary memory bottleneck. Theoretical analysis shows their gradients are decomposable into frame-independent terms. We therefore adopt an alternating computation-and-release gradient accumulation scheme: gradients are computed and accumulated per frame while immediately freeing intermediate variables before processing the next frame. This reduces memory complexity from $O(T)$ to $O(1)$, supporting stable optimization for long videos.

\subsubsection{Sparse-to-dense tracking information conversion}
\label{sssec:sparse_to_dense}

To balance system compatibility, editability, and computational efficiency, our method explicitly represents video spatial information using sparse tracking points. To recover per-pixel motion from this sparse representation, a motion propagation mechanism based on spatial smoothness priors is employed.
Specifically, for any pixel $\mathbf{p}$ in frame $i$, its corresponding position in frame $t$, is given by

\vspace{-15pt}

\begin{equation}
\mathbf{T}(\mathbf{p}, i, t) = \mathbf{p} - \mathbf{N}(\mathbf{p}, i) + \mathbf{T}_R(\mathbf{N}(\mathbf{p}, i), t)
\end{equation}

\vspace{-10pt}

\noindent where $\mathbf{N}(\mathbf{p}, i)$ represents the nearest sample point to $\mathbf{p}$ in frame $i$, and $\mathbf{T}_R(\mathbf{s}, t)$ denotes the coordinates of tracked sample point $\mathbf{s}$ in frame $t$. This method efficiently estimates full-frame motion while preserving the advantages of sparse representation.

\section{Results} \label{sec:results}

This section presents a comprehensive evaluation of our proposed method through multiple experiments, including: comparisons with SOTA methods (Sec \ref{subsec:sota_comparisons}), experiments on longer videos (Sec~\ref{subsec:long_videos}) and 3D animation-to-sketch conversion  (Sec~\ref{subsec:3d_to_sketch}), vector animation generation and frame interpolation from text-to-video models (Sec~\ref{subsec:text_to_sketch}), multiple ablation studies (Sec~\ref{subsec:ablation_studies}). Performance (Sec~\ref{subsec:performance}) and limitations (Sec~\ref{subsec:limitations}) are discussed in the end. Since our method produces dynamic vector animation sequences whose motion effects are difficult to fully capture in static images, readers are referred to the supplementary materials for complete dynamic comparisons.

\subsection{Comparisons with SOTA methods} \label{subsec:sota_comparisons}


Our approach is compared with SOTA methods, including detection-based methods (Canny~\cite{canny2009computational} and HED~\cite{xie2015holistically}), image-based sketching (CLIPasso~\cite{vinker2022clipasso}), video-specific sketch generation (SketchVideo~\cite{zheng2024sketch} and Fang et al.~\cite{fang2025video}), \change{and a modified variant derived from LiveSketch’s MLP architecture~\cite{gal2024breathing}, namely LiveSketch-MLP.}

\change {The original LiveSketch takes the first frame of a pre-existing sketch animation and a text prompt as inputs~\cite {gal2024breathing}, which deviates from the unified setting of our experiment (video input, sketch animation output). To resolve this discrepancy, we adapt the core MLP architecture of LiveSketch, initialize its first frame with the output of the first-frame generation module in our approach, and integrate it with the proposed loss function to construct LiveSketch-MLP. This modification ensures compliance with the input-output protocol of our experiment while preserving the architectural essence of the original model.}

As shown in Figure~\ref{fig:comparison-diff-methon}, Canny and HED produce results that resemble edge maps rather than semantically abstract sketches. While CLIPasso generates strokes with reasonable abstraction, its results exhibit poor temporal consistency, leading to noticeable visual flickering across frames. SketchVideo achieves better temporal coherence, but still suffers from inter-frame stroke jittering, particularly in regions with complex dynamic structures. VideoSketch proposed by Fang et al. demonstrates improved flicker suppression, yet shows instability in detailed areas (e.g., the flamingo's tail in Figure~\ref{fig:comparison-diff-methon}). Note that VideoSketch was applied to the entire image; only the foreground is shown for comparison purposes. \change{LiveSketch-MLP suffers from inter-frame jittering and exhibits semantic inconsistency, failing to capture the bent posture of the flamingo's head. In contrast, our method produces animation sequences without visible flickering across all frames, demonstrating superior temporal coherence and semantic consistency.}

Figure~\ref{fig:comparison-diff-pathnum} further compares results with varying numbers of strokes. Both CLIPasso and SketchVideo suffer from significant semantic loss with a small number of strokes (e.g., 4 strokes) and tend to fail to capture the original image content. \change{LiveSketch-MLP exhibits semantic inconsistency, especially when the frame index $t$ is large (the bottom line).} Our method maintains clear semantic expression and stable visual content even with a small number of strokes, validating the advantage of our DMT representation in cross-frame semantic modeling.

\begin{figure}[ht]
    \centering
    \includegraphics[width=\linewidth]{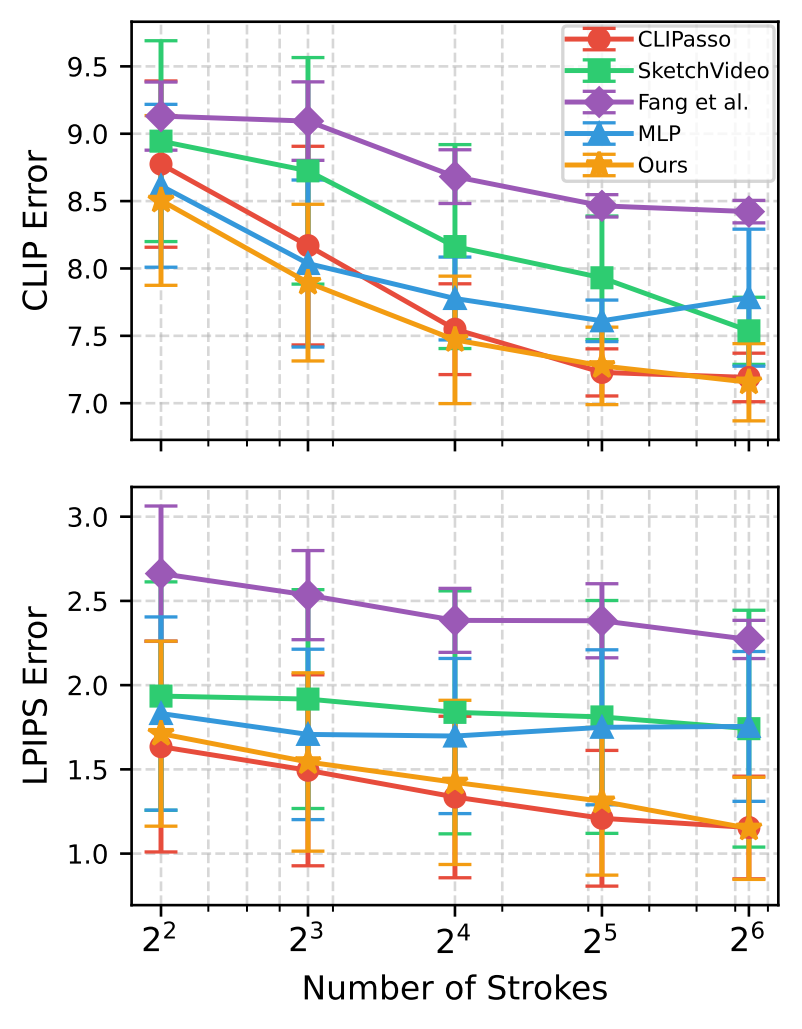} 


    \caption{The statistics of semantic error evaluation.}
    \label{fig:quantitative_results}
\end{figure}

We quantitatively evaluated semantic consistency via LPIPS~\cite{zhang2018unreasonable} and CLIP similarity~\cite{radford2021learning}, \change{with mean values and variances presented in Figure~\ref{fig:quantitative_results}}. Our method achieves comparable LPIPS scores to CLIPasso and outperforms SketchVideo\change{, Fang et al., and LiveSketch-MLP}; in the CLIP score, it is comparable to CLIPasso with more strokes and superior with fewer. However, since CLIPasso ignores temporal coherence and exhibits significant flickering, our approach has a clear overall advantage.

\subsection{Experiments on longer videos} \label{subsec:long_videos}


We conducted experiments on longer video sequences. Figure~\ref{fig:long_video_results} shows \change{the results of the surfing video with 300 frames and 6FPS from the LVOS dataset~\cite{hong2025lvos}. Additionally, experimental results of another 400-frame video \cite{yinliuzhizhu2025} are presented in Figure~\ref{fig:long_video_results_appendix} of the appendix.} Existing methods struggle to process such long sequences effectively, often failing due to computational resource constraints. In contrast, our approach achieves stable generation for such long videos by leveraging explicit sparse tracking point trajectories to guide stroke consistency, thus addressing a key limitation of previous methods. It is worth noting that benefiting from the continuity of DMT, our method provides a continuous temporal representation, enabling flexible adjustment of output frame rates, as shown in the surfing result.


\begin{figure}[t]
    \centering
        \includegraphics[width=0.32\linewidth]{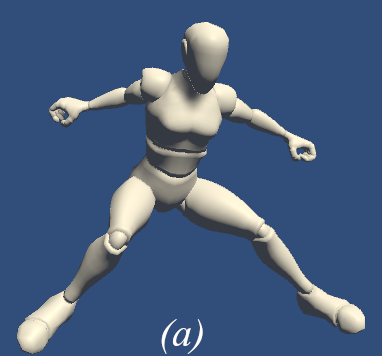} 
    \hfill
        \includegraphics[width=0.32\linewidth]{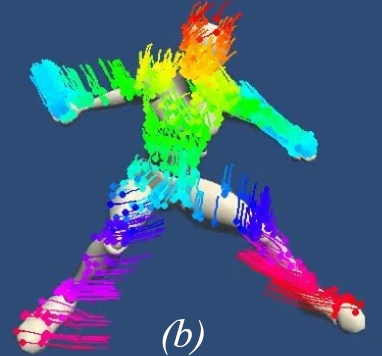} 
    \hfill
        \includegraphics[width=0.32\linewidth]{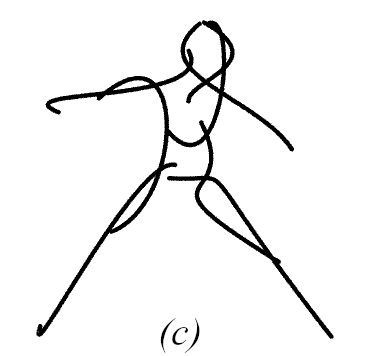} 
    \caption{Our result that converts a 3D animation to a sketch animation. Tracking information (b) is recorded during the 3D animation rendering (a), and (c) is one sketch result at this frame.}
    \label{fig:3d_results}
\end{figure}

\subsection{3D animation to sketch animation} \label{subsec:3d_to_sketch}

To validate the compatibility of our method with 3D animation data, we established a data capture pipeline using the Unity engine. We used a skinned character model~\cite{rpg_animations} with skeletal animations~\cite{human_motions}, using custom scripts to synchronize screen capture with sparse vertex trajectory recording. The experiment collected 500 frames while maintaining temporal consistency. Figure~\ref{fig:3d_results} shows such an example, taking the 3D animation as input.

\begin{figure}[t]
    \centering
    \begin{subfigure}[b]{\linewidth}
        \centering




        \vspace{0.05cm} 

\includegraphics[width=\linewidth]{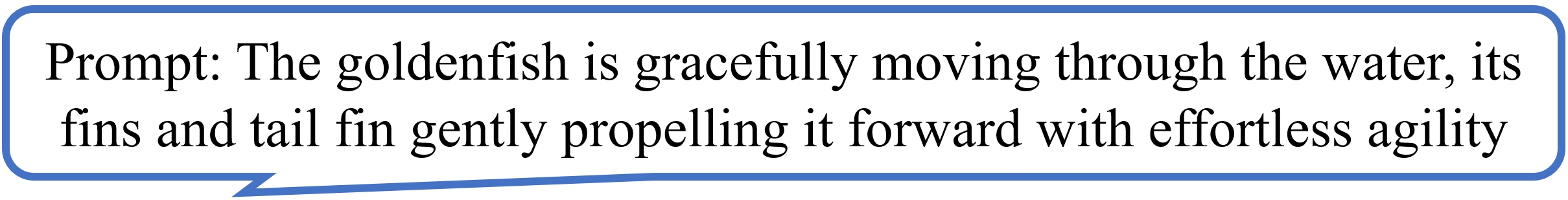} \\
        \includegraphics[width=0.15\linewidth]{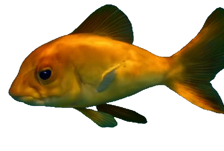} \hfill
        \includegraphics[width=0.15\linewidth]{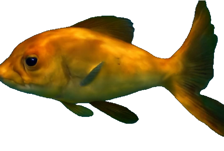} \hfill
        \includegraphics[width=0.15\linewidth]{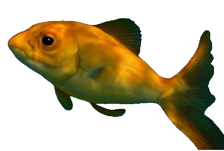} \hfill
        \includegraphics[width=0.15\linewidth]{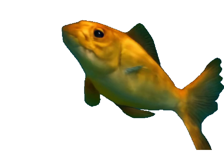} \hfill
        \includegraphics[width=0.15\linewidth]{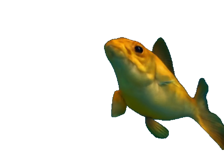}


        \includegraphics[width=0.15\linewidth]{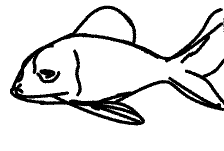} \hfill
        \includegraphics[width=0.15\linewidth]{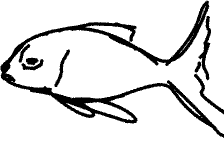} \hfill
        \includegraphics[width=0.15\linewidth]{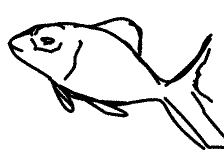} \hfill
        \includegraphics[width=0.15\linewidth]{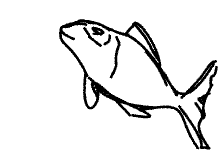} \hfill
        \includegraphics[width=0.15\linewidth]{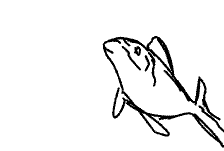}

    \end{subfigure}

    \caption{Our result of text-driven sketch animation generation. The input prompt, the generated video, and the vector sketch animation are shown from top to bottom.}
    \label{fig:text2video_results}
\end{figure}

\begin{figure*}[t]
    \centering
    \begin{tabular}{cc|cccc|cccc}    
        \multirow{3}{*}{
            \begin{tikzpicture}
            [scale=1, baseline=(current bounding box.center)]
            \draw[->, thick, line width=1pt] (0,1) -- (0,-1);
            \node at (0,-1.3) {$t$};
        \end{tikzpicture}}
        &
        \includegraphics[width=0.083\linewidth]{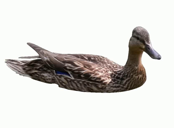} &
        \includegraphics[width=0.083\linewidth]{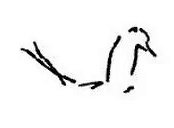} &
        \includegraphics[width=0.083\linewidth]{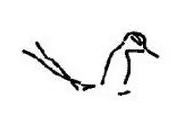} &
        \includegraphics[width=0.083\linewidth]{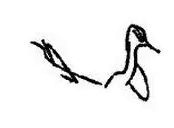} &
        \includegraphics[width=0.083\linewidth]{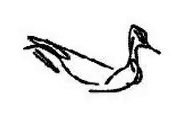} &
        \includegraphics[width=0.083\linewidth]{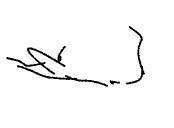} &
        \includegraphics[width=0.083\linewidth]{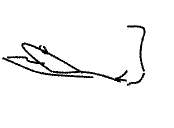} &
        \includegraphics[width=0.083\linewidth]{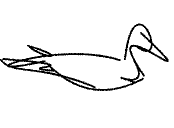} &
        \includegraphics[width=0.083\linewidth]{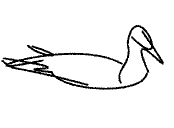} \\

        &
        \includegraphics[width=0.083\linewidth]{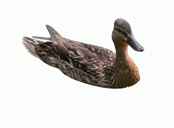} &
        \includegraphics[width=0.083\linewidth]{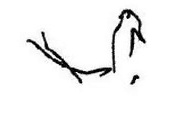} &
        \includegraphics[width=0.083\linewidth]{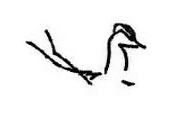} &
        \includegraphics[width=0.083\linewidth]{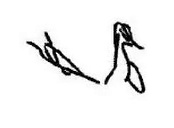} &
        \includegraphics[width=0.083\linewidth]{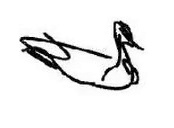} &
        \includegraphics[width=0.083\linewidth]{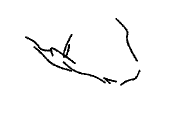} &
        \includegraphics[width=0.083\linewidth]{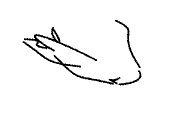} &
        \includegraphics[width=0.083\linewidth]{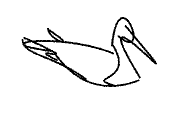} &
        \includegraphics[width=0.083\linewidth]{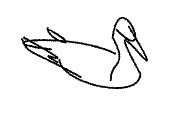} \\
        
        &
        \includegraphics[width=0.083\linewidth]{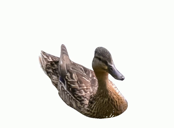} &
        \includegraphics[width=0.083\linewidth]{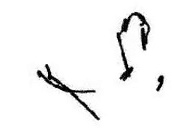} &
        \includegraphics[width=0.083\linewidth]{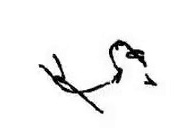} &
        \includegraphics[width=0.083\linewidth]{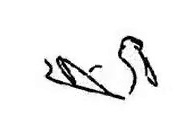} &
        \includegraphics[width=0.083\linewidth]{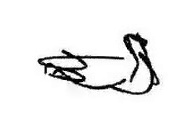} &
        \includegraphics[width=0.083\linewidth]{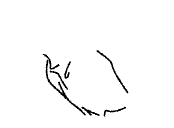} &
        \includegraphics[width=0.083\linewidth]{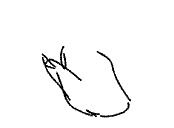} &
        \includegraphics[width=0.083\linewidth]{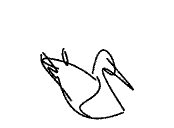} &
        \includegraphics[width=0.083\linewidth]{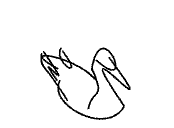} \\
          & \small{Input} & \small{50 iters} & \small{100 iters} & \small{500 iters} & \small{1000 iters} & 
        \small{50 iters} & \small{100 iters} & \small{500 iters} & \small{1000 iters} \\
        & \multicolumn{1}{c|}{} & \multicolumn{4}{c|}{Power basis} & \multicolumn{4}{c}{Bernstein basis} \\
    \end{tabular}
    \caption{Comparisons between power basis and Bernstein basis representations of DMT at 3 frames. The power basis shows slower convergence and distortions when $t$ increases, while the Bernstein basis converges faster and avoids distortion artifacts in all frames.}
    \label{fig:basis_comparison}
\end{figure*}

\subsection{Text to sketch animation} \label{subsec:text_to_sketch}

We explored text-to-sketch animation generation. We first employ CogVideoX~\cite{yang2024cogvideox} (the official CogVideoX-2B checkpoint) to generate an initial video from input text, then extract and crop the foreground object using a U2Net network~\cite{qin2020u2}, and finally convert the foreground video into vector sketch animation using our approach. Unlike methods that require additional sketch input~\cite{gal2024breathing,liu2025multi,liang2025multi}, our approach is based solely on text prompts, as shown in Figure~\ref{fig:text2video_results}. Note that the frame interpolation capability can mitigate the framerate limitations of existing text-to-video models.


\subsection{Ablation studies} \label{subsec:ablation_studies}

\subsubsection{Power basis vs. Bernstein basis} \label{subsubsec:basis_comparison}

We conducted ablation studies comparing power basis and Bernstein basis representations for our DMT module. As shown in Figure~\ref{fig:basis_comparison}, DMT using a power basis exhibits slower convergence, requiring more than 1,000 iterations to achieve reasonable results. Furthermore, when the parameter $t$ approaches larger values (i.e., posterior frames), the power-base representation tends to produce excessive distortions, which can not be solved with more iterations. In contrast, the Bernstein basis representation delivers faster convergence, achieving comparable quality within 500 iterations, and maintains distortion-free results for all frames, demonstrating superior numerical stability and optimization robustness.

\subsubsection{Motion heatmaps in initialization} \label{subsubsec:motion_heatmaps}

\begin{figure*}[t]
    \centering
        \includegraphics[width=0.135\linewidth]{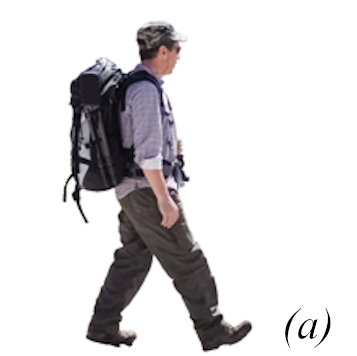} 
        \includegraphics[width=0.135\linewidth]{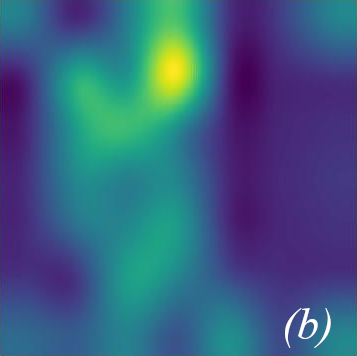} 
        \includegraphics[width=0.135\linewidth]{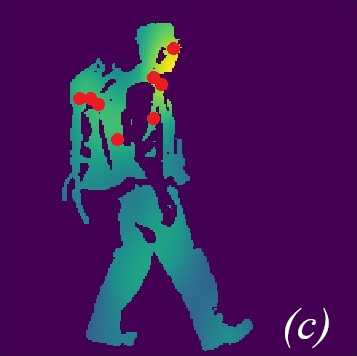} 
        \includegraphics[width=0.135\linewidth]{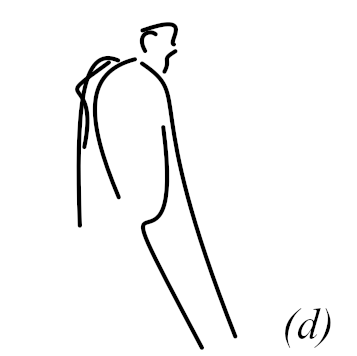} 
        \includegraphics[width=0.135\linewidth]{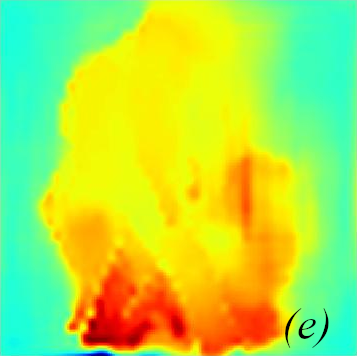} 
        \includegraphics[width=0.135\linewidth]{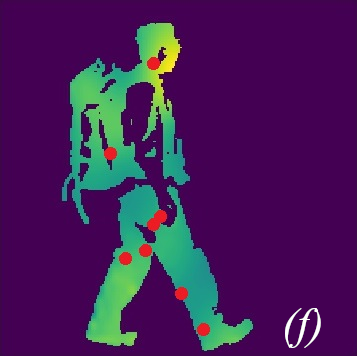} 
        \includegraphics[width=0.135\linewidth]{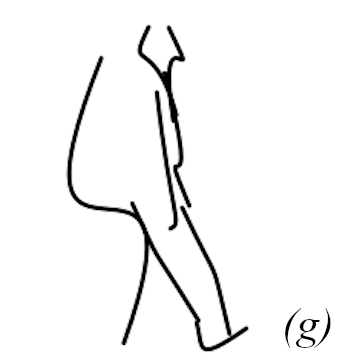} 
    \caption{Comparisons of stroke initialization strategies. (a) First frame of input video; (b) CLIP attention map; (c) Probability density map and sampled points without motion heatmap - insufficient sampling in the leg region; (d) Sketch result without motion heatmap that shows missing leg strokes; (e) Motion heatmap; (f) Probability density map integrated with motion heatmap - increased sampling in the leg region; (g) Our sketch result with motion heatmap, achieving a better depiction of structure.}
    \label{fig:heatmap_comparison}
\end{figure*}

We analyzed the role of motion heatmaps in stroke initialization through ablation studies. Figure~\ref{fig:heatmap_comparison} shows the impact on the probability distribution of stroke sampling. Without motion heatmaps, regions with frequent motion (e.g., legs) receive low probability density, leading to insufficient sampling when the number of strokes is strictly limited. Incorporating motion heatmaps significantly increases probability density in these regions, enabling better structural depiction even with a small number of strokes.

\subsubsection{Loss functions} \label{subsubsec:loss_functions}

\begin{figure}[t]
    \begin{subfigure}[a]{\linewidth}
        \centering
        \includegraphics[width=0.18\linewidth]{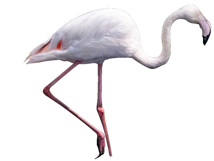}\hfill
        \includegraphics[width=0.18\linewidth]{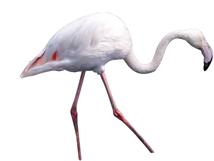}\hfill
        \includegraphics[width=0.18\linewidth]{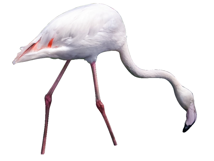}\hfill
        \includegraphics[width=0.18\linewidth]{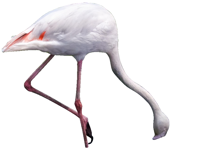}\hfill
        \includegraphics[width=0.18\linewidth]{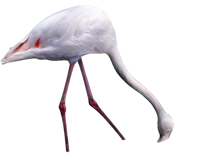} 
        \caption{Input}
    \end{subfigure}
    \begin{subfigure}[b]{\linewidth}
        \centering
        \includegraphics[width=0.18\linewidth]{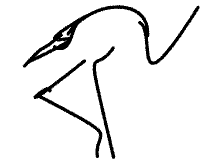}\hfill
        \includegraphics[width=0.18\linewidth]{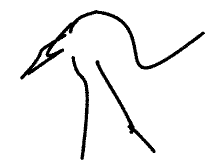}\hfill
        \includegraphics[width=0.18\linewidth]{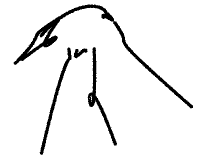}\hfill
        \includegraphics[width=0.18\linewidth]{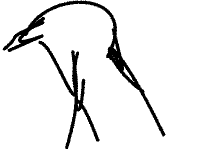}\hfill
        \includegraphics[width=0.18\linewidth]{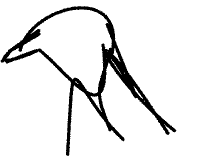}
        
        \caption{Results without consistency loss}
    \end{subfigure}
    \begin{subfigure}[c]{\linewidth}
        \centering
        \includegraphics[width=0.18\linewidth]{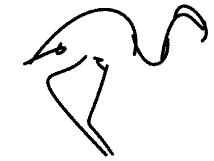}\hfill
        \includegraphics[width=0.18\linewidth]{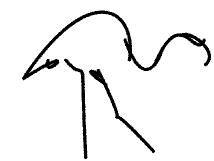}\hfill
        \includegraphics[width=0.18\linewidth]{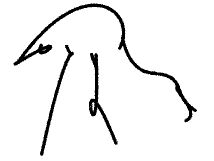}\hfill
        \includegraphics[width=0.18\linewidth]{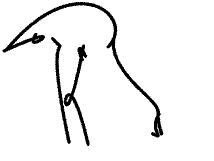}\hfill
        \includegraphics[width=0.18\linewidth]{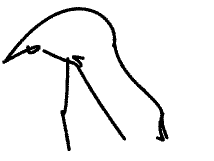}
        
        \caption{Results with consistency loss}
    \end{subfigure}
    
    \caption{Effect of consistency loss. Without consistency losses, strokes exhibit semantic misalignment across frames (e.g., leg strokes are confused with abdomen). The consistency loss maintains proper semantic correspondence.}
    \label{fig:consistency_ablation}
\end{figure}

We evaluated the impact of the proposed temporal consistency loss on the quality of sketch animation results. Without consistency loss, strokes often correspond to different semantic regions across frames, causing severe temporal inconsistency, including semantic confusion and unnatural structural distortions. As shown in Figure~\ref{fig:consistency_ablation}, flamingo leg strokes may incorrectly align with abdominal regions when consistency constraints are absent. Incorporating the temporal consistency loss maintains proper semantic correspondence across frames, effectively mitigating these issues.

\subsection{Performance} \label{subsec:performance}

\begin{table}[ht]
    \centering
    \small
    \begin{tabular}{lrr}
        \toprule
        \textbf{Method} &\textbf{frame} & \textbf{GPU Memory}\\
        \midrule
        SketchVideo & 50 & 17.2 GB\\
        \change{LiveSketch-MLP} & \change{50} & \change{4.58 GB}\\
        Ours & 50 / 400 & \textbf{2.47 GB / 3.37 GB}\\ 
         \midrule
        CoTracker (orig.)&50 & 11.38 GB\\
        CoTracker (opt.)&50 / 400 & \textbf{5.31 GB / 11.79 GB}\\            
        \bottomrule
    \end{tabular}
    \caption{GPU memory consumption is reduced significantly for both sketch animation generation and tracking information computation, allowing our approach to support long videos.}
    \label{tab:performance}
\end{table}

\change{Although the generated sketch animations are lightweight vector graphics, the generation methods based on pre-trained models (e.g., CLIP, StableDiffusion) are memory-intensive.} We evaluated performance on a workstation with an AMD EPYC 7B13 CPU and an NVIDIA GeForce RTX 3090 GPU. In the preprocessing stage, our memory optimization for CoTracker reduces its GPU memory consumption from 11.38~GB to 5.31~GB for a 50-frame video. As shown in Table~\ref{tab:performance}, the proposed generation method further demonstrates significant memory efficiency, requiring only 2.47~GB for 50 frames—an 85\% reduction over SketchVideo \change{and a 46\% reduction over MLP}. Even for long sequences (400 frames), our method maintains high memory efficiency, consuming only 3.37~GB. This enables execution on consumer-grade hardware (e.g., laptops with GTX 1050 Ti Mobile GPU), which previous methods could not support.

\begin{figure}[ht]
    \centering    
    \begin{tikzpicture}
        \begin{axis}[
            width=0.9\linewidth,
            height=0.4\linewidth,
            xlabel={}, 
            ylabel={Semantic consistency},
            xmode=log,
            log basis x=2,
            ymin=0.4, ymax=1, 
            grid=major,
            ylabel near ticks,  
            legend style={fill opacity=0.5, draw opacity=0.5, text opacity=1,font=\small,at={(1,1.75)}, anchor=north east}
        ]
        
        \addplot[blue, thick, mark=*] table {chart_data/user_sem_ours_vs_SketchVideo.data};
        \addplot[red, thick, mark=square*] table {chart_data/user_sem_ours_vs_CLIPasso.data};
        \addplot[green!70!black, thick, mark=triangle*] table {chart_data/user_sem_SketchVideo_vs_CLIPasso.data};
        \draw[dashed, red] (axis cs:0,0.5) -- (axis cs:65,0.5);
        \node at (axis cs:65,0.5) {50\%};
        \legend{Ours vs. SketchVideo, Ours vs. CLIPasso, SketchVideo vs. CLIPasso}
        \end{axis}
    \end{tikzpicture}
    
    \begin{tikzpicture}
        \begin{axis}[
            width=0.9\linewidth,
            height=0.4\linewidth,
            ylabel={Visual coherence},
            xmode=log,
            log basis x=2,
            ymin=0.4, ymax=1, 
            ylabel near ticks,  
            grid=major,
            legend style={fill opacity=0.5, draw opacity=0.5, text opacity=1,font=\small,at={(0.98,0.68)}, anchor=north east}
        ]
        
        \addplot[blue, thick, mark=*] table {chart_data/user_vis_ours_vs_SketchVideo.data};
        \addplot[red, thick, mark=square*] table {chart_data/user_vis_ours_vs_CLIPasso.data};
        \addplot[green!70!black, thick, mark=triangle*] table {chart_data/user_vis_SketchVideo_vs_CLIPasso.data};
        \draw[dashed, red] (axis cs:0,0.5) -- (axis cs:65,0.5);
        \node at (axis cs:65,0.5) {50\%};
        
        \end{axis}
    \end{tikzpicture}
    
    \begin{tikzpicture}
        \begin{axis}[
            width=0.9\linewidth,
            height=0.4\linewidth,
            xlabel={Number of strokes}, %
            ylabel={Artistic aesthetics},
            xmode=log,
            log basis x=2,
            ymin=0.4, ymax=1, 
            ylabel near ticks,  
            grid=major,
            legend style={fill opacity=0.5, draw opacity=0.5, text opacity=1,font=\small,at={(0.98,0.68)}, anchor=north east}
        ]
        
        \addplot[blue, thick, mark=*] table {chart_data/user_art_ours_vs_SketchVideo.data};
        \addplot[red, thick, mark=square*] table {chart_data/user_art_ours_vs_CLIPasso.data};
        \addplot[green!70!black, thick, mark=triangle*] table {chart_data/user_art_SketchVideo_vs_CLIPasso.data};
        \draw[dashed, red] (axis cs:0,0.5) -- (axis cs:65,0.5);
        \node at (axis cs:65,0.5) {50\%};
        
        \end{axis}
    \end{tikzpicture}

    \caption{\change{Pairwise comparison user studies: preference proportion for the former option in forced choices.}}
    
    \label{fig:user_study}
\end{figure}

\subsection{User studies} \label{subsec:user_study}
We recruited 20 volunteers to participate in a pairwise comparison user study, comparing our method, CLIPasso~\cite{vinker2022clipasso}, and SketchVideo~\cite{zheng2024sketch}. For each input, we displayed the input video alongside output videos produced by two methods in a randomized order. Participants were asked to select the result they preferred in three terms: semantic consistency, temporal coherence, and artistic aesthetics. The statistics, summarized in Figure~\ref{fig:user_study}, demonstrate an evident advantage for our method on all three criteria. Notably, this advantage was more pronounced in scenarios with a small number of strokes, highlighting our method's efficacy in creating concise and expressive sketch animations.

\subsection{Limitations} \label{subsec:limitations}

Our method exhibits several limitations that point toward valuable future research directions: 1) Our consistency loss relies on tracking information, which may be inaccurate in scenarios with rapid movements or severe occlusion. 2) Our iterative optimization process is computationally expensive; integrating diffusion models may improve efficiency~\cite{arar2025swiftsketch}. 3) Our approach focuses solely on foreground objects; one future direction would be to extend to full-screen videos\cite{vinker2023clipascene,fang2025video}.




\section{Conclusion} \label{sec:conclusion}


We present an automatic generation approach for vector sketch animation based on a novel Differentiable Motion Trajectory (DMT), which represents the frame-wise movement of stroke control points using differentiable polynomial-based trajectories. DMT enables global semantic gradient propagation across multiple frames, significantly improving the semantic consistency and temporal coherence, and is compatible with long video inputs and produces high-framerate output. Using a Bernstein basis, DMT can be stably optimized by balancing the sensitivity of polynomial parameters. 
Evaluations on DAVIS and LVOS datasets demonstrate the superiority of our approach over SOTA methods for both short and long videos. 
The robustness and compatibility of the proposed method are further validated through various ablation studies and successful applications in cross-domain scenarios, such as text-to-video generation and 3D animations.
Future research directions could be enhancing generation efficiency, increasing tracking robustness, and extending to various artistic and multimedia applications.




\section{\change{Acknowledgments}} \label{sec:acknowledgments}
\change{We would like to express our sincere gratitude to Prof.Xiaonan Fang for generously providing the experimental data of the method\cite{fang2025video}. We also acknowledge the authors of CLIPasso\cite{vinker2022clipasso} and CoTracker\cite{karaev2024cotracker3} for making their code publicly available. This work is supported by the National Natural Science Foundation of China (Grant No. 62172367).}



\appendix

\begin{table*}[]
    \begin{tabular}{l|l|lll|lll}
        \toprule
        \multirow{2}{*}{frames} & \multirow{2}{*}{Max degree} & \multicolumn{3}{c|}{Mean absolute error} & \multicolumn{3}{c}{Avg of abs values of coefficients} \\
         &  & Frame Sampling & Least Squares & Ridge Regression & Frame Sampling & Least Squares & Ridge Regression \\
         \hline
        50 & 24 & $2.849\cdot 10^2$ & \textbf{$5.687\cdot 10^{-2}$} & $7.782\cdot 10^{-2}$ & $1.588\cdot 10^9$ & $1.510\cdot 10^7$ & \textbf{$1.744\cdot 10^4$} \\
        100 & 49 & $1.485\cdot 10^{14}$ & \textbf{$6.293\cdot 10^{-2}$} & $1.695\cdot 10^{-1}$ & $1.182\cdot 10^{18}$ & $1.699\cdot 10^{13}$ & \textbf{$1.930\cdot 10^4$} \\
        200 & 99 & $5.857\cdot 10^{42}$ & \textbf{$8.988\cdot 10^{-2}$} & $5.231\cdot 10^{-1}$ & $6.264\cdot 10^{44}$ & $3.425\cdot 10^{13}$ & \textbf{$1.583\cdot 10^4$} \\
        400 & 199 & $8.320\cdot 10^{127}$ & \textbf{$1.727\cdot 10^{-1}$} & $1.433$ & $8.608\cdot 10^{129}$ & $2.936\cdot 10^{13}$ & \textbf{$2.004\cdot 10^4$}\\
        \bottomrule
    \end{tabular}
    \caption{Comparison of trajectory fitting methods on complex motion video. Ridge regression provides the best trade-off between accuracy and stability.}
    \label{tab:fitting_comparison}
\end{table*}

\begin{figure*}[]
    \centering
    \begin{subfigure}[b]{\textwidth}
        \centering
        
        \includegraphics[width=0.1\textwidth]{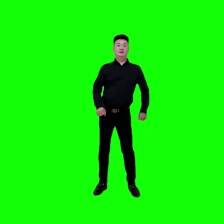}\hfill
        \includegraphics[width=0.1\textwidth]{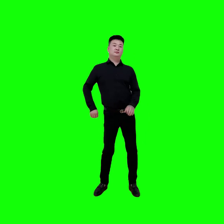}\hfill
        \includegraphics[width=0.1\textwidth]{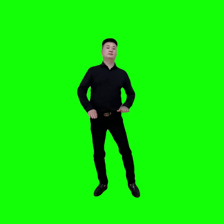}\hfill
        \includegraphics[width=0.1\textwidth]{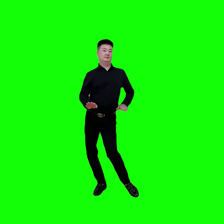}\hfill
        \includegraphics[width=0.1\textwidth]{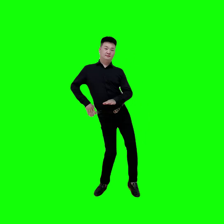}\hfill
        \includegraphics[width=0.1\textwidth]{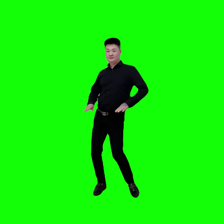}\hfill
        \includegraphics[width=0.1\textwidth]{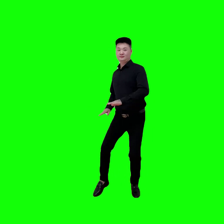}\hfill
        \includegraphics[width=0.1\textwidth]{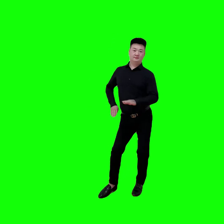}\hfill
        \includegraphics[width=0.1\textwidth]{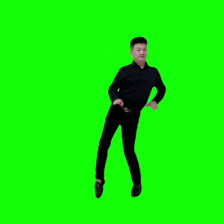}
        
        \stackunder{\includegraphics[width=0.1\textwidth]{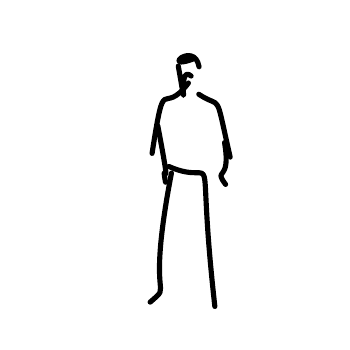}}{\small \# 8}\hfill
        \stackunder{\includegraphics[width=0.1\textwidth]{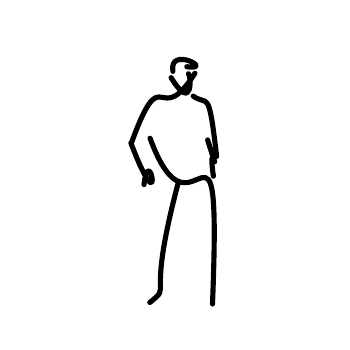}}{\small \# 18}\hfill
        \stackunder{\includegraphics[width=0.1\textwidth]{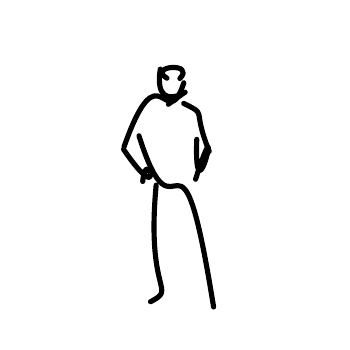}}{\small \# 28}\hfill
        \stackunder{\includegraphics[width=0.1\textwidth]{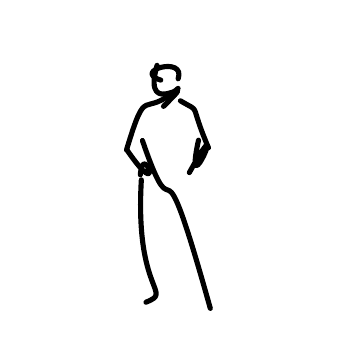}}{\small \# 38}\hfill
        \stackunder{\includegraphics[width=0.1\textwidth]{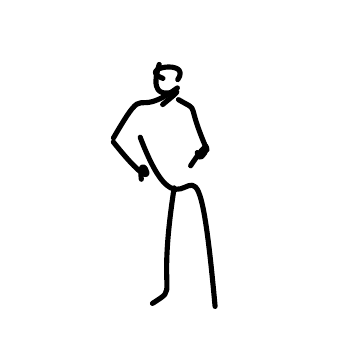}}{\small \# 48}\hfill
        \stackunder{\includegraphics[width=0.1\textwidth]{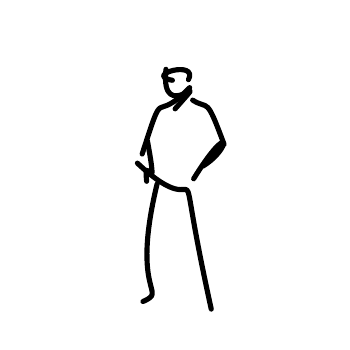}}{\small \# 58}\hfill
        \stackunder{\includegraphics[width=0.1\textwidth]{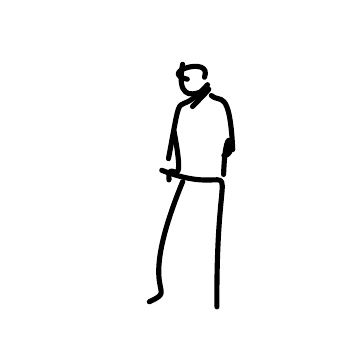}}{\small \# 68}\hfill
        \stackunder{\includegraphics[width=0.1\textwidth]{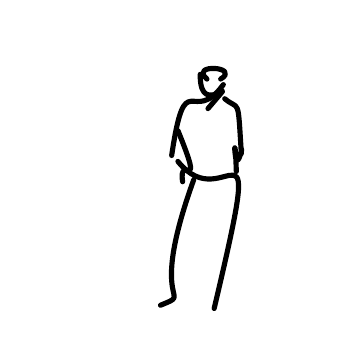}}{\small \# 78}\hfill
        \stackunder{\includegraphics[width=0.1\textwidth]{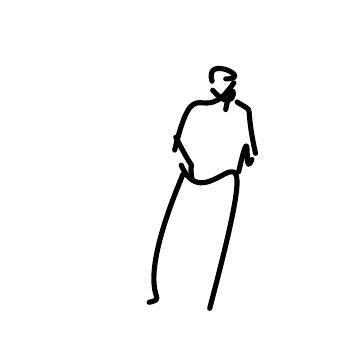}}{\small \# 88}

        \includegraphics[width=0.1\textwidth]{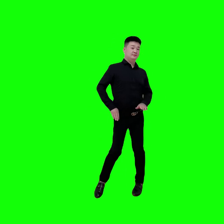}\hfill
        \includegraphics[width=0.1\textwidth]{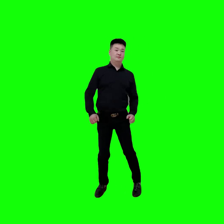}\hfill
        \includegraphics[width=0.1\textwidth]{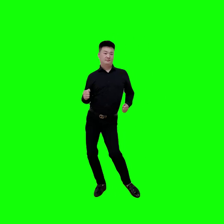}\hfill
        \includegraphics[width=0.1\textwidth]{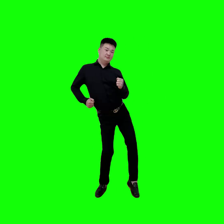}\hfill
        \includegraphics[width=0.1\textwidth]{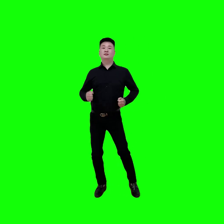}\hfill
        \includegraphics[width=0.1\textwidth]{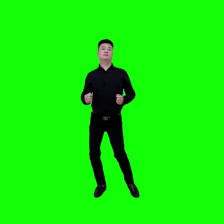}\hfill
        \includegraphics[width=0.1\textwidth]{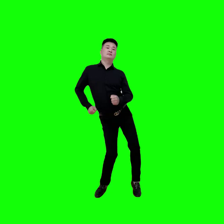}\hfill
        \includegraphics[width=0.1\textwidth]{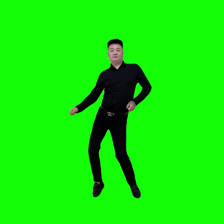}\hfill
        \includegraphics[width=0.1\textwidth]{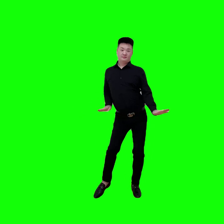}
        
        \stackunder{\includegraphics[width=0.1\textwidth]{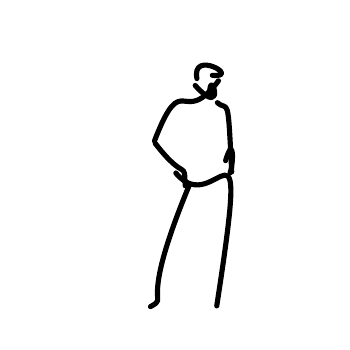}}{\small \# 98}\hfill
        \stackunder{\includegraphics[width=0.1\textwidth]{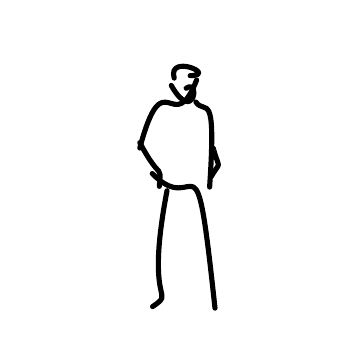}}{\small \# 108}\hfill
        \stackunder{\includegraphics[width=0.1\textwidth]{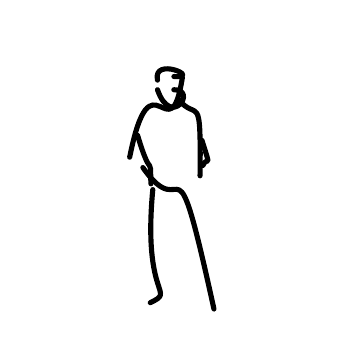}}{\small \# 118}\hfill
        \stackunder{\includegraphics[width=0.1\textwidth]{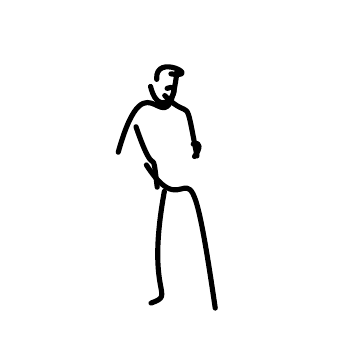}}{\small \# 128}\hfill
        \stackunder{\includegraphics[width=0.1\textwidth]{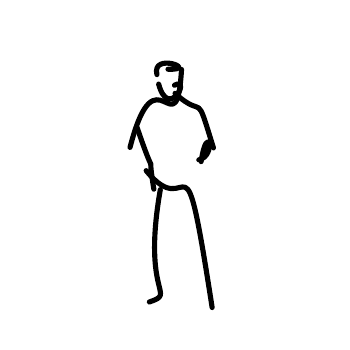}}{\small \# 138}\hfill
        \stackunder{\includegraphics[width=0.1\textwidth]{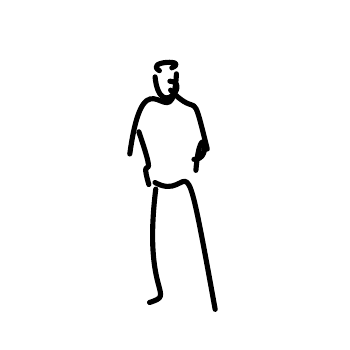}}{\small \# 148}\hfill
        \stackunder{\includegraphics[width=0.1\textwidth]{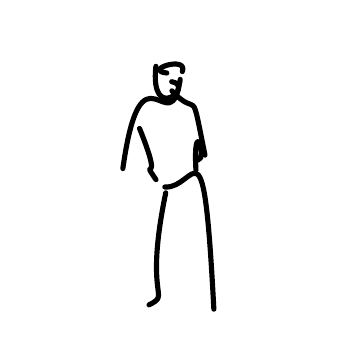}}{\small \# 158}\hfill
        \stackunder{\includegraphics[width=0.1\textwidth]{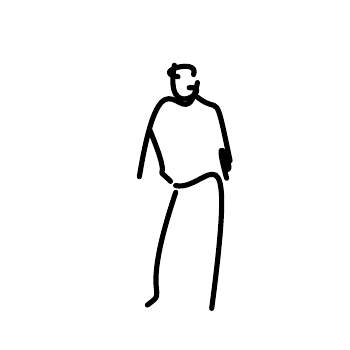}}{\small \# 168}\hfill
        \stackunder{\includegraphics[width=0.1\textwidth]{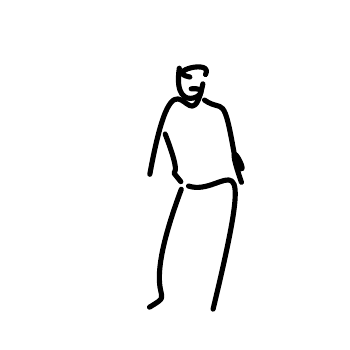}}{\small \# 178}

        \includegraphics[width=0.1\textwidth]{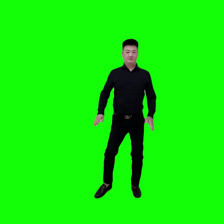}\hfill
        \includegraphics[width=0.1\textwidth]{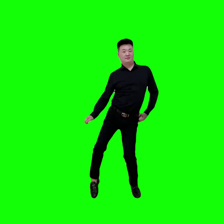}\hfill
        \includegraphics[width=0.1\textwidth]{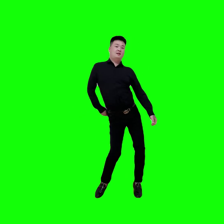}\hfill
        \includegraphics[width=0.1\textwidth]{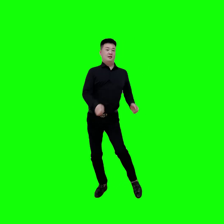}\hfill
        \includegraphics[width=0.1\textwidth]{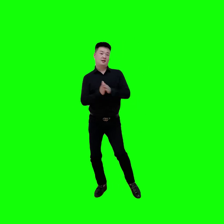}\hfill
        \includegraphics[width=0.1\textwidth]{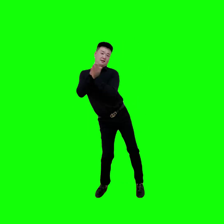}\hfill
        \includegraphics[width=0.1\textwidth]{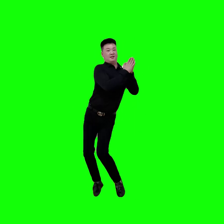}\hfill
        \includegraphics[width=0.1\textwidth]{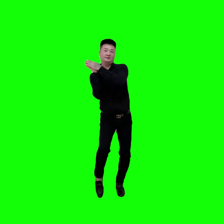}\hfill
        \includegraphics[width=0.1\textwidth]{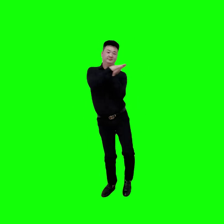}
        
        \stackunder{\includegraphics[width=0.1\textwidth]{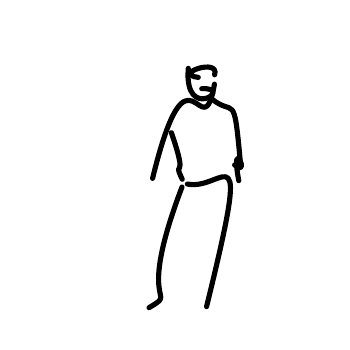}}{\small \# 188}\hfill
        \stackunder{\includegraphics[width=0.1\textwidth]{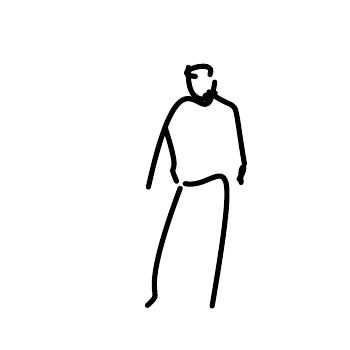}}{\small \# 198}\hfill
        \stackunder{\includegraphics[width=0.1\textwidth]{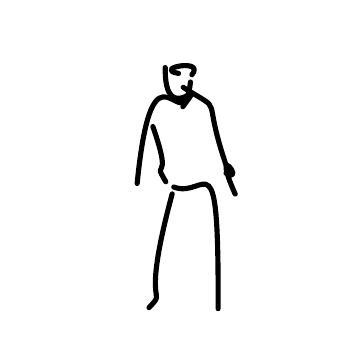}}{\small \# 208}\hfill
        \stackunder{\includegraphics[width=0.1\textwidth]{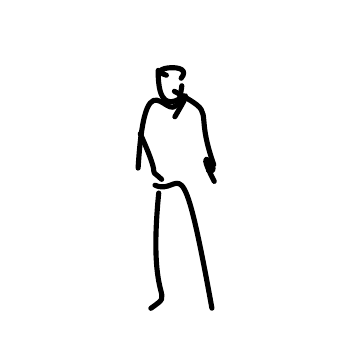}}{\small \# 218}\hfill
        \stackunder{\includegraphics[width=0.1\textwidth]{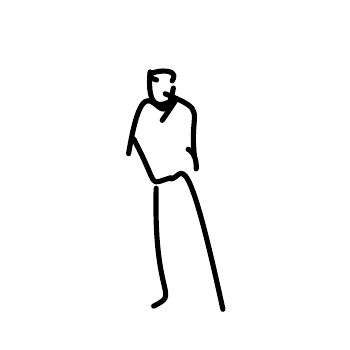}}{\small \# 228}\hfill
        \stackunder{\includegraphics[width=0.1\textwidth]{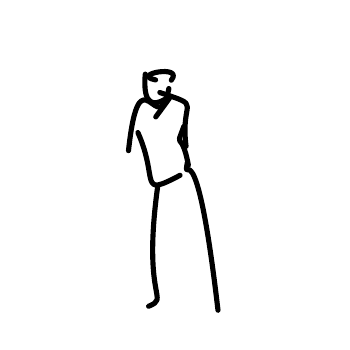}}{\small \# 238}\hfill
        \stackunder{\includegraphics[width=0.1\textwidth]{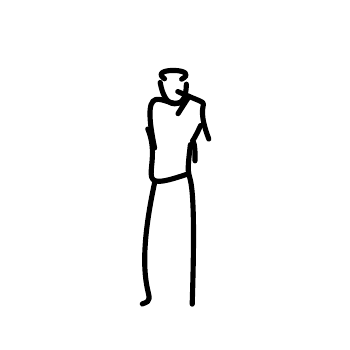}}{\small \# 248}\hfill
        \stackunder{\includegraphics[width=0.1\textwidth]{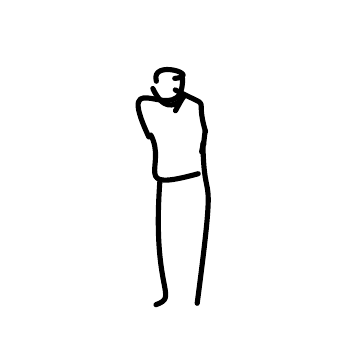}}{\small \# 258}\hfill
        \stackunder{\includegraphics[width=0.1\textwidth]{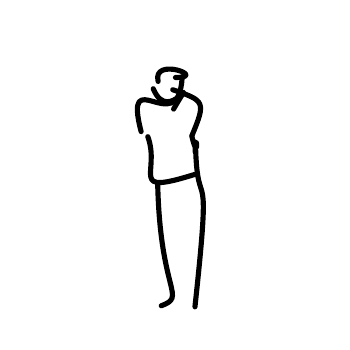}}{\small \# 268}

        \includegraphics[width=0.1\textwidth]{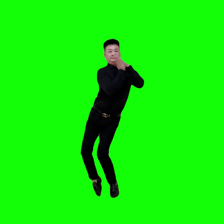}\hfill
        \includegraphics[width=0.1\textwidth]{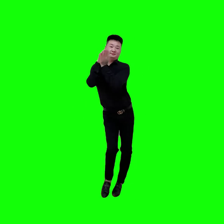}\hfill
        \includegraphics[width=0.1\textwidth]{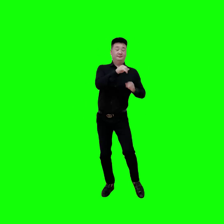}\hfill
        \includegraphics[width=0.1\textwidth]{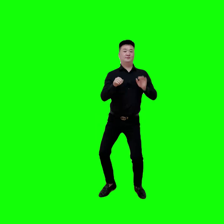}\hfill
        \includegraphics[width=0.1\textwidth]{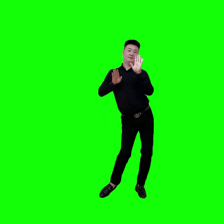}\hfill
        \includegraphics[width=0.1\textwidth]{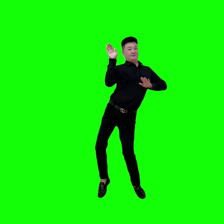}\hfill
        \includegraphics[width=0.1\textwidth]{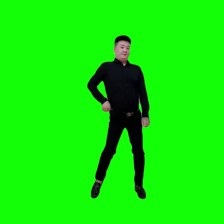}\hfill
        \includegraphics[width=0.1\textwidth]{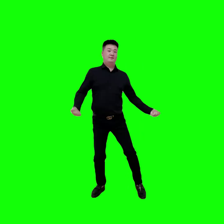}\hfill
        \includegraphics[width=0.1\textwidth]{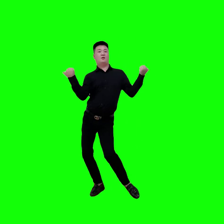}
        
        \stackunder{\includegraphics[width=0.1\textwidth]{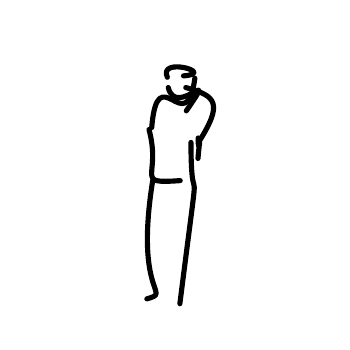}}{\small \# 278}\hfill
        \stackunder{\includegraphics[width=0.1\textwidth]{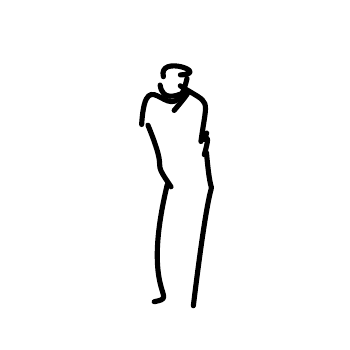}}{\small \# 288}\hfill
        \stackunder{\includegraphics[width=0.1\textwidth]{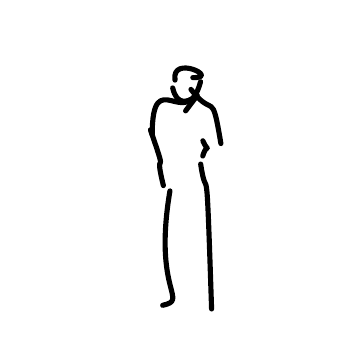}}{\small \# 298}\hfill
        \stackunder{\includegraphics[width=0.1\textwidth]{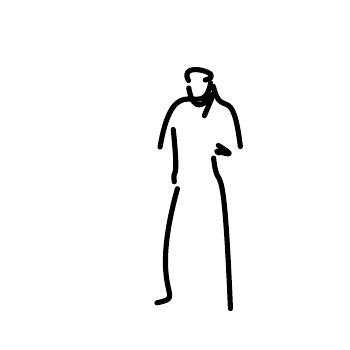}}{\small \# 308}\hfill
        \stackunder{\includegraphics[width=0.1\textwidth]{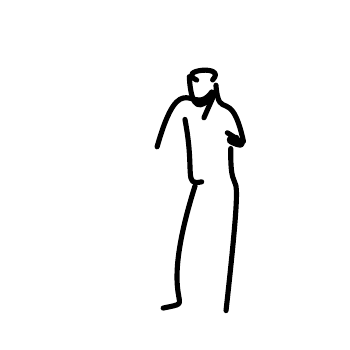}}{\small \# 318}\hfill
        \stackunder{\includegraphics[width=0.1\textwidth]{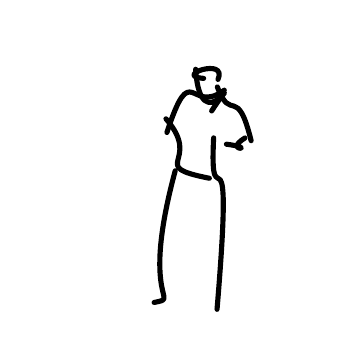}}{\small \# 328}\hfill
        \stackunder{\includegraphics[width=0.1\textwidth]{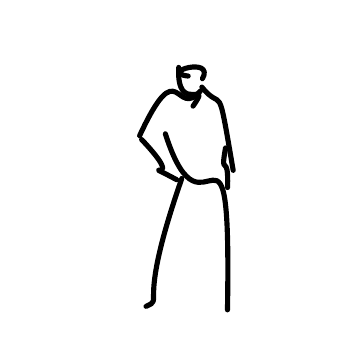}}{\small \# 338}\hfill
        \stackunder{\includegraphics[width=0.1\textwidth]{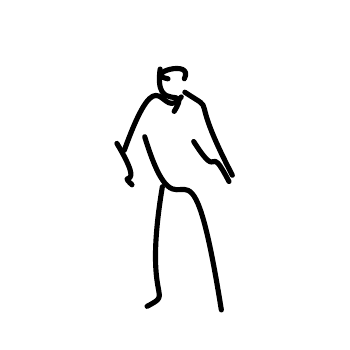}}{\small \# 348}\hfill
        \stackunder{\includegraphics[width=0.1\textwidth]{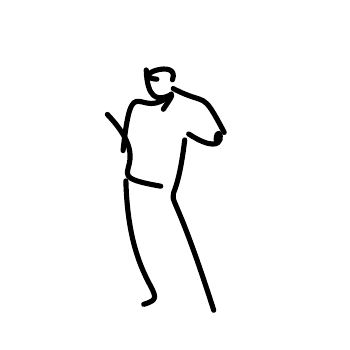}}{\small \# 358}

    \end{subfigure}
    
    \caption{\change{The result of \textbf{long Dance video} with 400 frames}~\cite{yinliuzhizhu2025}}
    \label{fig:long_video_results_appendix}
\end{figure*}


\appendix

\section{Feasibility Proof of Polynomial Approximation} \label{app:feasibility_proof}

This appendix provides the detailed mathematical proof supporting the feasibility of representing control point motion using polynomial functions.

\noindent\textbf{Theorem 1.} If a Bézier curve changes continuously over time, then its control points must also vary continuously.

\begin{proof}
Assuming that the Bézier curve changes continuously over time, let the curve at time \(t\) be \(C(u,t)\). Since the Bernstein basis functions are linearly independent, the control points \(\{P_0, P_1, ..., P_m\}\) can be uniquely determined from the curve \(C(u)\). 

For a set of distinct values \(u\), e.g. \(u_0, u_1, ..., u_m\), if the curve \(C(u)\) is continuous with respect to \(t\) for each \(u\), then the corresponding points on the curve \(C(u_0), C(u_1), ..., C(u_m)\) are also continuous with respect to \(t\). These points on the curve are linear combinations of the control points, which we can express as:
\[
\begin{bmatrix}
C(u_0, t) \\
C(u_1, t) \\
\vdots \\
C(u_m, t)
\end{bmatrix}
=
\begin{bmatrix}
B_{m,0}(u_0) & B_{m,1}(u_0) & \cdots & B_{m,m}(u_0) \\
B_{m,0}(u_1) & B_{m,1}(u_1) & \cdots & B_{m,m}(u_1) \\
\vdots & \vdots & \ddots & \vdots \\
B_{m,0}(u_m) & B_{m,1}(u_m) & \cdots & B_{m,m}(u_m)
\end{bmatrix}
\cdot
\begin{bmatrix}
P_0(t) \\
P_1(t) \\
\vdots \\
P_m(t)
\end{bmatrix}
\]

Abbreviated as:
\[
\mathbf{C}(t) = \mathbf{M} \cdot \mathbf{P}(t)
\]

Here, \(\mathbf{M}\) is an invertible matrix (when the \(u_k\) are chosen appropriately), allowing us to solve for the control points:
\[
\mathbf{P}(t) = \mathbf{M}^{-1} \cdot \mathbf{C}(t)
\]

Therefore, if the Bézier curve is continuous in \(t\), its control points must also be continuous in \(t\).
\end{proof}

\noindent\textbf{Theorem 2.} A continuously changing Bézier curve can be arbitrarily approximated by a Bézier curve whose control points are polynomial functions of time \(t\).

\begin{proof}
For an arbitrary control point \(P\), its 2D coordinate vector at time parameter \(t \in [0,1]\) can be represented by a function \(P(t)\), which is continuous in \(t\). According to the Weierstrass approximation theorem~\cite{stone1948generalized}, any continuous function can be arbitrarily approximated by polynomials. Thus, for each control point \(P_i(t)\), there exists a polynomial function \(Q_i(t)\) (with time \(t\) as the variable) such that \(|P_i(t) - Q_i(t)| < \epsilon\) holds for all \(t \in [0,1]\), where \(\epsilon\) is an arbitrarily small positive number.

Define the approximate Bézier curve represented by \(Q_i(t)\) as \(D(u,t) = \sum_{i=0}^m B_{m,i}(u) Q_i(t)\). Since the Bernstein basis satisfies \(\sum_{i=0}^m B_{m,i}(u) = 1\) and is non-negative, for any \(u, t \in [0,1]\), we have:

\begin{align*}
|C(u,t) - D(u,t)| 
&= \left| \sum_{i=0}^m (P_i(t) - Q_i(t)) B_{m,i}(u) \right| \\
&\leq \sum_{i=0}^m |P_i(t) - Q_i(t)| B_{m,i}(u) \\
&\leq \epsilon \sum_{i=0}^m B_{m,i}(u) = \epsilon.
\end{align*}

This means the error between the approximate curve \(D(u,t)\) and the original curve \(C(u,t)\) does not exceed \(\epsilon\) for any \(u, t \in [0,1]\). Therefore, a continuously changing Bézier curve can be arbitrarily approximated by a Bézier curve whose control points are polynomial functions of time \(t\).
\end{proof}


\section{Stroke trajectory initialization} 
\label{sec:appendix}

We compare three fitting strategies for initializing stroke control point motion trajectories: uniform sampling, least squares, and ridge regression.

\noindent \textbf{polynomial interpolation} selects $n+1$ frames from the video, uses normalized time $t_i$ as nodes and solves linear equations to ensure the trajectory passes exactly through sample points in these frames. Though fitting is accurate at sampled frames, fitting error increases significantly at non-sampled frames when $n > 15$.

\noindent \textbf{Least squares} minimizes the sum of squared errors for polynomial coefficient estimation. However, when polynomial degree is high ($n > 40$), the Vandermonde matrix becomes ill-conditioned~\cite{bell1978solutions}, making solutions highly sensitive to noise and numerically unstable.

\noindent \textbf{Ridge regression} introduces L2 regularization~\cite{hoerl1970ridge} to address ill-conditioning in high-degree polynomial fitting ($n > 200$). Although Mean Absolute Error (MAE) is similar between least squares and ridge regression, the former produces coefficients with extremely large magnitudes, leading to oscillatory behavior and overfitting. Ridge regression constrains coefficient magnitudes, yielding stable and smooth trajectories.

Table~\ref{tab:fitting_comparison} compares the three methods on the complex dance video under varying frame counts and polynomial degrees. Frame sampling shows large errors and coefficient magnitudes. Least squares achieves lower MAE but exhibits unstable trajectories due to large coefficients. Ridge regression maintains acceptable MAE (maximum 1.433 pixels) while keeping coefficient magnitudes around $10^4$, demonstrating significantly better stability.

\bibliographystyle{alpha-doi}
\bibliography{ref}

\end{document}